\documentclass[a4paper,fleqn]{cas-dc}

\usepackage{algorithm}
\usepackage{algpseudocode}
\usepackage[numbers]{natbib}

\usepackage{amsmath,amssymb,bm}
\usepackage{physics}
\usepackage{siunitx}

\usepackage{tikz}
\usepackage{amsmath}
\usepackage{bm}
\usetikzlibrary{arrows.meta, calc, patterns}

\usepackage{float}
\usepackage{placeins}

\def\tsc#1{\csdef{#1}{\textsc{\lowercase{#1}}\xspace}}
\tsc{WGM}
\tsc{QE}
\tsc{EP}
\tsc{PMS}
\tsc{BEC}
\tsc{DE}

\ExplSyntaxOn
\cs_set:Npn \__first_footerline:
{
  \group_begin:
  \small
  \sffamily
  \ifnum\theblind>0\relax
  \else
	\__short_authors:
  \fi
  \group_end:
}
\ExplSyntaxOff

\begin{document}
\let\WriteBookmarks\relax
\def\floatpagepagefraction{1}
\def\textpagefraction{.001}
\shorttitle{Unresolved DEM model for platelet dynamics in blood flow}
\shortauthors{Zucchelli F. et~al.}

\title [mode = title]{Modelling platelet dynamics in blood flow: an unresolved DEM approach}

\author[1]{Francesca Zucchelli}
\cormark[1]
\ead{francesca.zucchelli@unige.ch}

\affiliation[1]{organization={Computer Science Department, University of Geneva},
                addressline={1227 Carouge},
                country={Switzerland}}

\author[2]{Carmine Porcaro}
\author[2]{Mahdi Saeedipour}

\affiliation[2]{organization={Department of Particulate Flow Modelling, Johannes Kepler University},
                addressline={A-4040 Linz}, 
                country={Austria}}

\author[1]{Bastien Chopard}
\author[1]{Jonas Latt}

\cortext[cor1]{Corresponding author}
\begin{abstract}
Computational models of blood flow are caught between fully resolved cell-based methods, which faithfully reproduce the dynamics of individual cells but are computationally prohibitive at vessel scale, and continuum models, scalable yet blind to particle motion. We present an unresolved, mesoscale computational fluid dynamics-discrete element method (CFD-DEM) model of platelet dynamics that bridges this gap and, coupled with an existing unresolved description of deformable red blood cells, moves closer to a scalable model of whole blood. Within this framework, implemented on the open source OpenFOAM-LIGGGHTS coupling, platelets are represented as rigid oblate particles advanced by orientation dependent drag, lift and hydrodynamic torque closures. The model is first validated against resolved simulations and experiments in cylindrical microvessels of diameter $100-200\,\mu m$, across wall shear rates $\dot{\gamma}=150-1650\,s^{-1}$ and hematocrit Ht $=10-20\%$ then used to characterize platelet margination. The model quantifies diffusion coefficient, CFL formation and their dependence on shear rate, Ht and channel size. In particular, platelet diffusivity grows with vessel size and shear rate, while remaining insensitive to Ht, whereas CFL thickens with shear rate and thins with Ht. We further show that even at this mesoscale level, the model remains sensitive to platelet shape: oblate platelets marginate faster and diffuse nearly an order of magnitude more than their spherical surrogates, reaching a comparable steady state distribution but along markedly different temporal paths. Together these results demonstrate that shape driven platelet dynamics can be recovered at a fraction of the cost of fully resolved methods.
\end{abstract}


\maketitle

\section{Introduction}
Blood carries out some of the most fundamental tasks in our body. It is responsible for transport of nutrients, oxygen and hormones, regulation of temperature as well as protection against diseases and fluid loss in case of damages to the vasculature. It is a suspension of cells and cell fragments in liquid phase, the plasma. The most numerous cells are red blood cells (RBCs), they make up around 45\% of the physiological volume fraction. Less numerous but indispensable for the body immune response are white blood cells, and finally platelets (PLTs), with a number density of $150{,}000$ - $400{,}000\,\mu l^{-1}$ \cite{Walker1990}.\\
PLTs are anucleated cellular fragments with characteristic diameters of 2-$3\,\mu m$ and a discoid resting shape, much smaller than the deformable biconcave RBCs which are around $8\,\mu m$ in diameter \cite{Thon2012}. They play a key role in hemostasis and are highly involved in wound repair, constantly monitoring the integrity of the endothelium as they circulate through the cardiovascular system. In the event of vessel lesion, when the sub endothelial matrix is exposed to blood, PLTs rapidly initiate the coagulation cascade that culminates in the formation of a fibrinogen-stabilized platelet plug, in the end forming a proper clot \cite{Thon2012, VanHinsbergh2012}.\\
It is not by chance that PLTs reside so close to the vessel wall, it is simply the consequence of RBCs migration away from the walls. As red cells deform under arterial flow conditions they experience wall-induced lift forces that instigate their cross stream motion towards the vessel center, which in return empties the near-wall region of cells \cite{Turitto1975, Turitto1996}. It is this annular lubrication layer, often referred to as cell-free layer (CFL), that is responsible for the reduced apparent viscosity of blood, as per the notorious F\aa hr\oe us–Lindqvist effect \cite{Fahreus1931}. In the meantime, by the process called margination, PLTs are pushed out from the core region of vessels and drift to the CFL from which they hardly ever escape \cite{Aarts1988, TillesEckstein1987, Reasor2013}.\\
Margination of PLTs emerges from a combination of hydrodynamic lift, RBC--PLT collisions and near-wall interactions, and it is sensitive to platelet shape as proven by a recent computational study by Ye et al. on the lateral migration of sphere-like and oblate-like particles \cite{Ye2018, Chang2018, Reasor2013}. One of the key physical factors regulating the phenomenon is shear rate, as clearly evidenced by the early experimental studies of Tilles \& Eckstein \cite{TillesEckstein1987, Eckstein1988}. Their work shows how margination of platelet sized particles flowing in glass channels containing red cells, increases as shear rate in the channel increases, affecting at the same time the CFL size which grows as more and more PLTs reach the walls. Near-wall excess of particles though, does not increase monotonically with shear rate, instead it reaches a plateau which defines the optimal shear rate value that maximizes margination \cite{Kruger2016, Li2023}. Above this threshold, which changes with vessel diameter, both the flux of particles marginating and the CFL thickness looses its dependence on shear rate \cite{Freund2011, Li2023}. Hematocrit is another influencing parameter of PLTs near-wall excess, indeed margination occurs solely in the presence of RBCs as it was well described through the experiments by Turitto \& Goldsmith or the earlier studies by Aarts et al. as well as the more recent simulations of Ye et al. \cite{CorattiylEckstein1986, TillesEckstein1987, Eckstein1988, Turitto1975, Turitto1996, Aarts1988, Ye2018}. As Ht increases, the volume exclusion effect of RBCs on PLTs increases, therefore enhancing margination. The mechanism of margination is all but simple, indeed shear rate and Ht don't operate in isolation, but influence each other constantly. For example: at low shear rate, PLTs--RBCs collisions are more frequent the higher the Ht and as a consequence margination increases with RBC volume fraction; on the contrary, at higher shear rate the collision frequency decreases despite the Ht, negatively impacting margination times \cite{Dynar2024, Ye2018}.\\
When it comes to the computational modeling of blood, its multiscale nature, the diversity of cells involved, as well as the high number of interactions that continuously occur between these cells and their intrinsic complexity, make the task particularly challenging. Currently available blood flow models present limitations that require the user to make a crucial choice: continuum-based models or cell-based simulations. The former allow for large-scale simulations and longer time scales, but cannot capture particle dynamics, indeed RBCs and PLTs, or any other cellular species, are considered in terms of their concentrations in plasma. One interesting example is the work by Sorensen et al. who employ a series of convection-diffusion-reaction equations to numerically study thrombus formation and growth \cite{Sorensen1999a,Sorensen1999b}, or the study by Cardillo and Barakat who adopted a similar approach to develop a model able to account for shear gradient when simulating platelet plug formation \cite{Cardillo2025}. Cell-based simulations on the other hand, can represent RBC deformability and PLTs motion with high fidelity (e.g. Immersed Boundary Method (IBM), Dissipative Particle Dynamics (DPD), Coarse-Grained Molecular Dynamics (CGMD) are all methods vastly employed in particle-based blood modeling), however they are computationally prohibitive when the domain size extends to vessel networks or when the time scale reaches seconds of physical time, both of which are required for clinically relevant scenarios. For example, Crow and Fogelson used a Lattice Boltzmann-IBM (LB-IBM) in two-dimensional 50 $\mu m$ wide channels to find how platelet diffusivity varies with closeness to the channel walls \cite{CrowlFogelson2010, CrowlFogelson2011}. A Lattice Boltzmann-Finite Element method was also applied by Kotsalos et. al to study PLT diffusin in shear flow \cite{Kotsalos2022}, whereas Yazdani and Karniadakis employed a DPD method to characterize the influence of shear rate and hematocrit on margination in constricted 30 $\mu m$ diameter channels \cite{Yazdani2016}. The same method was used by Fedosov et al. to study the development of the CFL and its thickness under different Ht\% and flow conditions \cite{Fedosov2010}. Zàvodsky et al. employed the LB-IBM solver HemoCell to quantify RBC and PLT diffusivity across a wide range of flow conditions, linking PLT margination not to gradients in diffusivity but to the cross stream gradient of Ht\% itself \cite{Zavodszky2019}, instead Peng Zhang et al. adopted CGMD to develop a model that accurately reproduces platelet molecular constituents and their cytoskeletral biomechanical properties in blood flow \cite{Zhang2017}.\\
Unresolved CFD-DEM provides an alternative, a compromise between fidelity and scalability, allowing large-scale microfluidic simulations with hundreds of thousands of particles to be carried out with efficiency \cite{Kloss2012}, a measure of which is given at the end of Section \ref{sec:setup}. In this paradigm, plasma is modeled as a continuum fluid solved on an Eulerian mesh, whereas RBCs and PLTs are represented as Lagrangian particles whose deformation and interaction are captured through effective mechanical models. Moreover, hydrodynamic coupling is achieved via closure laws for sub-grid forces, producing a model that is orders of magnitude cheaper than fully resolved methods while retaining the ability to simulate dense suspensions and capture emergent phenomena such as the F{\aa}hr{\ae}us--Lindqvist effect and cell margination. This mesoscale approach was recently adopted by Porcaro et al. to propose an unresolved blood model where deformable RBC dynamic is recovered from the upscaling of fully resolved LBM-IBM simulations \cite{Kotsalos2019, PorcaroSaeedipour2024, PorcaroSaeedipour2025}.\\
The present work builds on that RBC-focused framework and adds platelets, the missing component needed to construct a model of whole unresolved blood. The dynamics of PLTs in suspension was recovered from parametrical models describing drag and lift forces acting specifically on ellipsoidal particles and then implemented to fit the existing computational architecture \cite{Ouchene2020}. This way, the upscaled platelets respect the anisotropy, orientation and torque-driven rotation specific to their cellular geometry, yielding a more physiologically realistic behavior than a simplistic spherical approximation, as corroborated by our results. Finally, to explore the ability of the model to reproduce platelet susceptibility to ambient parameters, shear rate and Ht\% were varied in a set of numerical validations demonstrating their effects on margination and particle diffusion in a cylindrical microvessel with RBCs.\\
The article is organized as follows: first the unresolved CFD--DEM method is thoroughly described together with the models underlying RBCs and PLTs dynamics, then results concerning the simulation of blood cells in channel flow are presented and discussed, finally concluding remarks are provided.

\section{Numerical Method}
Blood was modeled as a suspension of RBCs and PLTs in Newtonian incompressible plasma, the carrier fluid. The simulations presented in this work were performed within the framework of the open-source software CFDEMCoupling, combining OpenFOAM to handle the CFD part and LIGGGHTS for the DEM part. The next sub-sections will be dedicated to explaining the coupled computational method and the force models driving particle motion in fluid.     
\subsection{Unresolved CFD--DEM}
With CFDEMCoupling, the coupling between the fluid solver and the particle dynamics solver can occur at the particle scale or at the mesoscale, labeling the former algorithm \textit{resolved} CFD--DEM and the latter \textit{unresolved} CFD--DEM. In both cases, a Finite Volume method is used to simulate the fluid dynamic solving Navier-Stokes equations, and a Discrete Element Method is used to solve two equations for the conservation of particle linear and angular momentum. The difference between the two algorithms lies in the fluid mesh refinement to particle size ratio.

Being $\Delta x$ the size of a fluid  element and $d_p$ the diameter of the smallest particle in the system, \textit{resolved} CFD--DEM is such that $\frac{d_p}{\Delta x} > 1$. The computational mesh is finer than a particle diameter; hence each particle will take up multiple fluid cells, allowing to fully calculate at each timestep the force exerted by the fluid over the particle interface by simple space integration \cite{Goniva2012, PorcaroSaeedipour2024}. On the contrary, $\frac{d_p}{\Delta x} < 1$ in the case of \textit{unresolved} CFD--DEM, meaning the fluid mesh is coarser than the diameter of a particle, so each fluid cell can contain multiple particles. As a result, the more traditional form of the volume-averaged continuity and momentum equations are adjusted to account for the volume fraction $\phi$ of solid in each fluid cell. Assuming an incompressible fluid of constant density, the resulting modified Navier-Stokes equations are:
\begin{align}
  & \pdv{(1-\phi)}{t} + \nabla\cdot[(1-\phi)\bm{u}] = 0, \label{eq:continuity}\\
  & \rho\left\{\pdv{[(1-\phi)\bm{u}]}{t} + \nabla\cdot[(1-\phi)\bm{u}\bm{u}]\right\} = \notag\\
  & -(1-\phi)\nabla p + (1-\phi)\mu\nabla^2\bm{u} - \bm{R}_{p,f}. \label{eq:momentum}
\end{align}
Here, $\rho$ and $\mu$ are the plasma density and dynamic viscosity, while the term $\bm{R}_{p,f}$ accounts for the momentum exchange between fluid and particles (the equal and opposite to the sum of hydrodynamic forces acting on the particles contained in a fluid cell, distributed as a volumetric source term) \cite{PorcaroSaeedipour2024}. On the other hand, the DEM equations used to compute particles position and velocity need to account for a change in the force expressing the fluid-particle interaction:
\begin{align}
  & m_p \dv{\bm{v}_p}{t} = m_p\bm{g} + \bm{F}^f_p + \sum_{N_p}\bm{F}^{p}_{p} + \sum_{N_w}\bm{F}^{w}_{p} + \bm{F}^{\mathrm{ext}}_p, \label{eq:dem_trans}\\
  & \bm{I}_p \dv{\bm{\omega}_p}{t} = \bm{T}^{f}_p + \sum_{N_p}\bm{T}^{p}_{p} + \sum_{N_w}\bm{T}^{w}_{p} + \bm{T}^{\mathrm{ext}}_p, \label{eq:dem_rot}
\end{align}
with $m_p$ the particle mass, $\bm{I}_p$ its inertia tensor, and $\bm{F}^{f}_{p}$,$\bm{F}^{p}_{p}$, $\bm{F}^{w}_{p}$, $\bm{F}^{ext}_{p}$, $\bm{T}^{f}_{p}$, $\bm{T}^{p}_{p}$, $\bm{T}^{w}_{p}$, $\bm{T}^{ext}_{p}$ the particle-fluid, particle-particle, particle-wall, and partice-external source interaction forces and the corresponding torques. Here $\bm{F}^f_p$ and consequently $\bm{T}^{f}_p= \bm{r}\times\bm{F}^f_p$ cannot be numerically computed by integrating the flow conditions at the fluid-particle interface over the particle surface, they need to be modeled. First, the term $\bm{F}^f_p$ is split into four components:
\begin{align}
    \bm{F}^f_p = \bm{F}^f_{p,v} + \bm{F}^f_{p,p} + \bm{F}^f_{p,\mathrm{drag}} + \bm{F}^f_{p,\mathrm{lift}}
\end{align}
the viscous, pressure, drag and lift forces acting on the particle, respectively. $\bm{F}^f_{p,v}$ and $\bm{F}^f_{p,p}$ have been extensively discussed in \cite{Goniva2012}, while drag and lift forces, as well as torque, will be treated in the following sections for RBCs and PLTs respectively. 

\subsection{RBC drag and lift models}
The RBC component follows the unresolved DEM formulation introduced in previous work \cite{Kotsalos2019, PorcaroSaeedipour2024, PorcaroSaeedipour2025}, where deformability is represented by a coarse-grained viscoelastic model that captures the effective shear elasticity and volume constraints of the membrane at the particle level. For completeness, the RBC parametrization used in our simulations is summarized below.\\
RBCs were modeled as volume equivalent spherical particles with characteristic diameter $d_{\mathrm{RBC}}\approx\SI{6}{\micro\meter}$, but retaining the dynamics of deformable biconcave cells. Two are the characteristic motions of RBCs: tank-treading happens in shear flows when the viscosity contrast between carrier fluid and cell content is low (the cell settles on a fixed orientation while its membrane rotates around the core fluid), tumbling on the other hand appears when the viscosity contrast is higher and is so that the cell rotates as a rigid body around its center of mass \cite{Dynar2024, PorcaroSaeedipour2024}. To ensure this physiological motion, highly accurate resolved data was employed to extrapolate drag and lift parametrical laws for RBCs \cite{PorcaroSaeedipour2024}. In further detail, the equations of the forcing terms read:
\begin{align}
    & \bm{F}^f_{RBC,drag} =0.125C_d\rho\pi d^2_{RBC} |\bm{u}- \bm{u}_p|(\bm{u}-\bm{u}_p),\label{eq:FD_rbc}\\
    & \bm{F}^f_{RBC,lift} = \frac{C_l\rho d^6_{RBC}|\bm{\nabla u} \cdot \bm{a}|^{2}(\nabla \bm{u} \cdot \bm{a})}{\nu},\label{eq:FL_rbc}
\end{align}
with $C_d$ and $C_l$ the drag and lift coefficients respectively, function of the particle Reynolds number $Re_p=\frac{d_{RBC}(u_p-u)}{\nu}$,
\begin{align}
    & C_d(Re_p)=\frac{73.146}{Re_p(1+29.86Re_p^{27.079})},\label{eq:CD_rbc}\\
    & C_l(Re_p)=\frac{100.271}{\left( \frac{\dot{\gamma}d_{RBC}^2}{\nu}\right)^{0.63}},\label{eq:CL_rbc}
\end{align}
with $\dot{\gamma}$ the plasma shear rate and $\nu$ the kinematic viscosity of plasma.

RBC--RBC and RBC--wall contacts are treated with a dissipative Hertz--Mindlin model, combining a nonlinear elastic normal force with velocity-dependent damping derived from the coefficient of restitution, and a tangential spring-damper friction force limited by the Coulomb threshold.

\subsection{PLTs drag, lift and torque models}
\label{sec:plt_rigidbody}
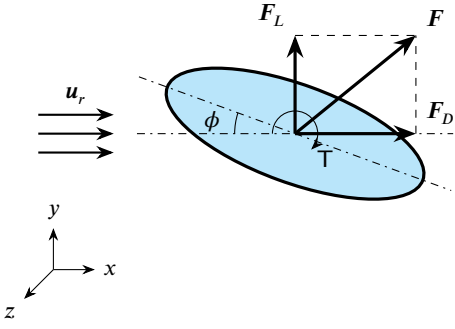
\begin{figure}
\centering
\begin{tikzpicture}[
  >=Stealth,
  thick,
  every node/.style={font=\small}
]
 
\def\lu{3.5}
\def\ld{4.0}
\def\hc{2.2}
 
\coordinate (O) at (0,0);
 
\def\elA{1.8}
\def\elB{0.65}
\def\teta{-20}
 
\begin{scope}[rotate=\teta]
  \filldraw[fill=cyan!25, draw=black, line width=1.2pt]
    (O) ellipse [x radius=\elA, y radius=\elB];
    \draw[dash dot, thin] (-\elA-0.4, 0) -- (\elA+0.4, 0);
\end{scope}

\draw[->, thin] (-0.3, 0) arc (180:-45:0.3);
\node at (0.4, -0.33) {T};

\draw[-, thin] (-0.8, 0) arc (180:162:0.89);
\node at (-1.1, 0.17) {$\phi$};
 
\draw[dash dot, thin] (-0.25\columnwidth, 0) -- (0.25\columnwidth, 0);

\begin{scope}[overlay]
\foreach \y in {0.25, 0, -0.25}{
  \draw[->, thick] (-3.4, \y) -- (-2.4, \y);
}
\node at (-2.9, 0.5) {$\bm{u}_r$};
\end{scope}
 
\def\FLx{0.0} \def\FLy{1.3}
\draw[->, line width=1.2pt] (O) -- (\FLx,\FLy) node[above left] {$\bm{F}_L$};
 
\def\FDx{1.6} \def\FDy{0.0}
\draw[->, line width=1.2pt] (O) -- (\FDx,\FDy) node[above right] {$\bm{F}_D$};
 
\draw[dashed, thin] (\FLx,\FLy) -- (\FDx+\FLx, \FDy+\FLy);
\draw[dashed, thin] (\FDx,\FDy) -- (\FDx+\FLx, \FDy+\FLy);
\draw[->, line width=1.2pt] (O) -- (\FDx+\FLx, \FDy+\FLy)
  node[above right] {$\bm{F}$};
 
\def\axLen{0.55}
\begin{scope}[overlay]
  \coordinate (axO) at (-\lu+0.3, -\hc+0.4);
  \draw[->, thin] (axO) -- ++(0, \axLen)  node[above]      {$y$};
  \draw[->, thin] (axO) -- ++(\axLen, 0)  node[right]      {$x$};
  \draw[->, thin] (axO) -- ++({-\axLen*0.707},{-\axLen*0.707}) node[below left] {$z$};
\end{scope}
 
\end{tikzpicture}
\vspace{1.5cm}
\caption{Schematics of forces and torque acting on an oblate particle as per the model of Ouchene \cite{Ouchene2020}. }
\label{fig:platelet_ouchene_moel}
\end{figure}
Unlike RBCs, platelets and their dynamics have not, to the authors' knowledge, been extensively studied as isolated particles in plasma. Resting platelets can however be reasonably approximated as rigid oblate particles \cite{Moskalensky2018}. Therefore, within this \textit{unresolved} framework, platelet motion was modeled using drag, lift and torque formulations developed for oblate ellipsoids in uniform flow \cite{Ouchene2020}. Consequently, the drag, lift and torque correlations used in this work are derived for an isolated oblate particle in unbounded uniform flow, as sketched in Figure \ref{fig:platelet_ouchene_moel}. In the \textit{unresolved} CFD--DEM framework, these relations are employed as local closures, assuming that at low Reynolds number the hydrodynamic forces are primarily determined by the instantaneous slip velocity and particle orientation. Under this assumption, a Poiseuille flow is locally approximated as uniform at the particle scale. It should be noted that this approximation neglects shear-induced and many-body hydrodynamic effects, as well as confinement corrections. However, the resulting model is shown to reproduce macroscopic platelet transport features such as margination, supporting its suitability at this level of modeling. As stated in the introduction, this paper aims to display that modeling platelets as oblate particles, rather than approximating their dynamics as spherical, leads to more physiologically realistic behavior, given the strong influence of particle geometry on flow patterns. The equations implemented as forcing terms were the following:
\begin{align}
 \|\bm{F}^f_{\mathrm{PLT},\mathrm{drag}}\| =
C_D \left(\frac{1}{2}\rho_f\|\bm{u}_r\|^2\frac{\pi}{4}d_{\mathrm{PLT}}^2\right)
,\label{eq:FD_plt}\\
  \|\bm{F}^f_{\mathrm{PLT},\mathrm{lift}}\| =
C_L \left(\frac{1}{2}\rho_f\|\bm{u}_r\|^2\frac{\pi}{4}d_{\mathrm{PLT}}^2\right), \label{eq:FL_plt}
\end{align}
with $C_D$ and $C_L$ the drag and lift coefficients respectively \cite{Ouchene2020}:
\begin{equation}
\left\{
\resizebox{\columnwidth}{!}{$
\begin{aligned}
& C_D = C_{D,\phi=0^\circ}
+ \left(C_{D,\phi=90^\circ}-C_{D,\phi=0^\circ}\right)\sin^2\phi,\\
& C_{D,\phi=0^\circ} =
\frac{24}{\mathrm{Re}}\left(
K_{\phi=0^\circ}
+ 0.15\,\lambda^{95.91}\mathrm{Re}^{0.687}
+ 0.2927(1-\lambda)^{0.4374}\mathrm{Re}^{0.7512}
\right),\\
& C_{D,\phi=90^\circ} =
\frac{24}{\mathrm{Re}}\left(
K_{\phi=90^\circ}
+ 0.15\,\lambda^{100.7}\mathrm{Re}^{0.687}
+ 0.1411(1-\lambda^{24.75})\mathrm{Re}^{0.7143}
\right),
\end{aligned}
$}
\right.
\label{eq:CD_plt}
\end{equation}
\begin{equation}
\resizebox{\columnwidth}{!}{$
C_L =
\left[
(C_{D,\phi=0^\circ}-C_{D,\phi=90^\circ})
+
\frac{2\left(\frac{1-\lambda}{\lambda}\right)^{0.542}}
{(1+\lambda)^{7.85}}
\mathrm{Re}_p^{0.1516}
\right]
\sin\phi \cos\phi.
$}
\label{eq:CL_plt}
\end{equation}
In contrast to spherical particles, oblate particles experience orientation-dependent hydrodynamic forces when subject to non-uniform flows. As a result, hydrodynamic torque must also be considered, since the particle orientation relative to local flow influences the particle angular velocity. For spherical particles this effect is absent, as their hydrodynamic response is independent of orientation. The employed torque model reads:
\begin{align}
    & ||\bm{T}^f_{PLT}|| = C_T \left(\frac{1}{2}\rho_f\|\bm{u}_r\|^2\frac{\pi}{8}d_{\mathrm{PLT}}^3\right),\label{eq:T_mag}
\end{align}
with $C_T$ the oblate torque coefficient computed as
\begin{align}
    & C_T = \frac{1.85\left({\frac{1-\lambda}{\lambda}}\right)^{0.832}}{Re_p^{0.146}}\sin\phi \cos\phi. \label{eq:CT_plt}
\end{align}

It is imperative here to reiterate that although platelets are considered as oblates in the force and torque calculations, the CFDEMCoupling particle representation retains a spherical placeholder (the diameter of the volume equivalent sphere being $d_{PLT}\approx2.66\mu m$). The particle orientation must still be tracked to compute the torque acting on the effective oblate shape and to achieve this, an orientation-tracking method was implemented to extend the open-source version of the solver without introducing a dedicated ellipsoidal particle type. This will be the focus of the next subsection.\\
These oblate force and torque models were compared against a very simplistic drag and lift formulation for spheres which was solely aimed at ensuring particle migration towards 60\% of the channel radius as this is the nature of particle radial migration in Poiseuille flow, which was first theorized by Segrè and Silberberg \cite{SegreSilberberg1961} and later confirmed within the scope of blood applications by Aarts et al. \cite{Aarts1988}. The corresponding drag force follows the Schiller-Naumann model, thus reading like equation (\ref{eq:FD_plt}), deviating only on the expression of the coefficient $C_D$:
\begin{align}
    & C_D = \frac{24}{Re}\left(1 + 0.15\,Re^{0.687}\right). \label{eq:CD_shiller}
\end{align}
Instead, the lift force was implemented as follows:
\begin{equation}
    \resizebox{\columnwidth}{!}{$
    \begin{aligned}
        & ||\bm{F}_{L,i}|| = 0.125 \frac{\rho ||\bm{u}_{\max}||^2 d_{PLT}^4}{D^2} \, C_{L}, \\
        & C_L = a \, \beta \left(\frac{D}{4||\bm{u}_{\max}||}\right)\dot{\gamma}_i
        + b \, Re^2 \left[\left(\frac{D}{4||\bm{u}_{\max}||}\right)\dot{\gamma}_i\right]^2
        + c \, \beta \left[\left(\frac{D}{4||\bm{u}_{\max}||}\right)\dot{\gamma}_i\right]^3, \label{eq:CL_shearlift}
    \end{aligned}
    $}
\end{equation}
    
with \textit{D} the channel diameter, $\beta$ confinement ratio, $a=110.88$, $b=0.00206$, $c=-302.345$ parameters. This semi-empirical shear induced lift model combines the works of McLaughlin et al., Mei et al. and Loth\&Dorgan \cite{Mclaughlin1991, Mei1992, Loth2009}.

\begin{figure*}
	\centering
	\includegraphics[width=.9\textwidth]{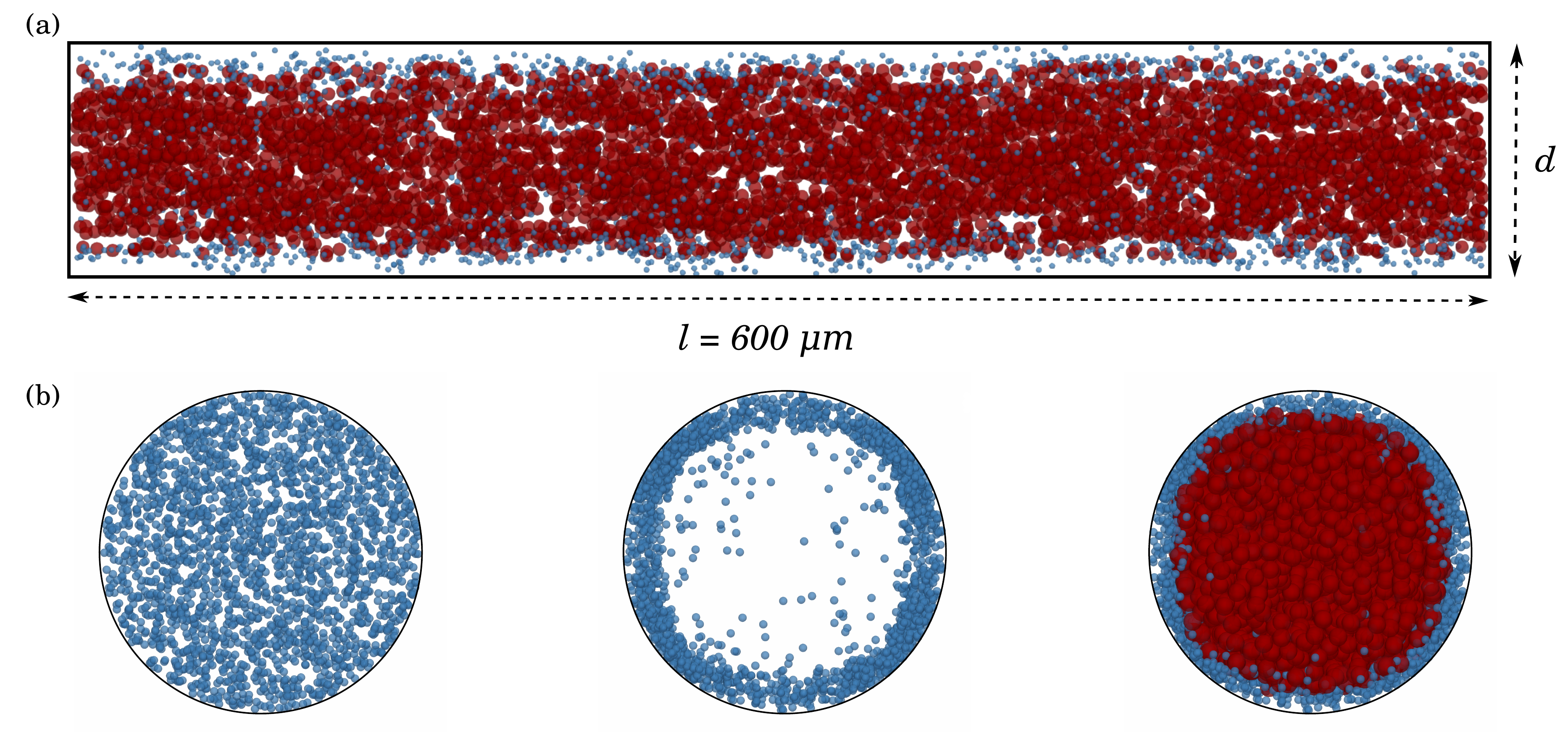}
	\caption{Example of simulation setup, with fixed length $l=600\,\mu m$ and varying diameter $d =[100,150,200]\,\mu m$. (a) Lateral view at steady state of a vessel, here $d = 100\,\mu m$, $Ht = 10\%$ and the CFL is clearly visible. (b) Vessel front views, from left to right, highlighting the PLTs initial condition, the steady state distribution of PLTs alone and combined.}
	\label{fig:setup}
\end{figure*}

\subsection{Implementation of particle orientation }
\label{subsec: orientation}
In the present work, each CFDEMCoupling sphere representing a PLT was analytically considered as an oblate particle of aspect ratio $AR=c/a\approx0.25$, with $a=2.06\,\mu m$ being one of the two long semi-axis and $c=0.52\,\mu m$ being the short semi-axis \cite{Mody2005}. The particle is treated as a rigid body with mass $m_{\mathrm{PLT}}=8.79\,pg$ and principal moments of inertia (in the body frame):
\begin{align}
  & I_{xx}=I_{yy}=\frac{m_{\mathrm{PLT}}}{5}\left(a^2+c^2\right),\qquad
  I_{zz}=\frac{2m_{\mathrm{PLT}}}{5}a^2
\end{align}
with principal axis $z$ aligned with the oblate symmetry axis $\bm{\vec{n}}$.\\
To avoid the singularities associated with Euler-angle para metrization, PLT orientation was represented by a unit quaternion $\textbf{q}=(q_w,q_x,q_y,q_z)$, where $\bm{q}$ defines the rotation from the particle body frame to the world frame. Details of the quaternion formulation can be found for example in \cite{ComputationalGranularDynamics2005}. The corresponding rotation matrix $\bm{R}(\bm{q})$ was used to transform vectors between frames. The particle angular velocity $\bm{\omega}$ and torque $\bm{T}^f_{PLT}$ were first mapped from the world frame to the body frame,
\begin{align}
  & \bm{\omega}_{b} = \bm{R}(\bm{q})^{\top}\bm{\omega}_{w}, \qquad
  \bm{T}^f_{PLT,b} = \bm{R}(\bm{q})^{\top}\bm{T}^f_{PLT,w}
\end{align}
and rotational dynamics were then advanced in the body frame using Euler's rigid-body equations,
\begin{align}
  \bm{I}_{b}\,\dot{\bm{\omega}}_{b}
  = \bm{T}^f_{PLT,b} - \bm{\omega}_{b}\times(\bm{I}_{b}\bm{\omega}_{b})
\end{align}
with $\bm{I}_{b}=\mathrm{diag}(I_x,I_y,I_z)$. After updating $\bm{\omega}_{b}$ explicitly, the angular velocity was transformed back to the world frame,
\begin{align}
  \bm{\omega}_{w} = \bm{R}(\bm{q})\,\bm{\omega}_{b}.
\end{align}
The orientation was then advanced using the world-frame angular velocity through
\begin{align}
  \dot{\bm{q}} = \frac{1}{2}\bm{\Omega}(\bm{\omega}_{w})\,\bm{q}, \\
  \bm{\Omega}(\bm{\omega}_{w})=
    \begin{bmatrix}
    0 & -\omega_x & -\omega_y & -\omega_z\\
    \omega_x & 0 & \omega_z & -\omega_y\\
    \omega_y & -\omega_z & 0 & \omega_x\\
    \omega_z & \omega_y & -\omega_x & 0
    \end{bmatrix},
\end{align}
followed by normalization of $\bm{q}$ to preserve unit norm and avoid numerical drift.\\
When it came to RBCs, they were always treated as spheres, so the anisotropic rigid-body update was bypassed and instead the angular velocity was advanced directly from the applied torque using the scalar moment of inertia $I=0.4\,m_{RBC}r^2$.\\
The steps implemented to handle time integration of particle orientation are summarized by Algorithm \ref{alg: orientation}.

\begin{algorithm}[H]
\caption{Quaternion-based rotational update with body-frame inertia integration}
\label{alg: orientation}
\begin{algorithmic}[1]
\For{each particle $i$}
  \State Advance translational velocity using applied force
  \State Advance particle position

  \If{particle $i$ uses anisotropic inertia}
    \State Compute $\bm R(\bm q_i)$ from the current quaternion
    \State Map angular velocity $\bm{\omega}$ and torque $\bm{T}^f_{PLT}$ to the body frame:
    \State Compute body-frame angular momentum $\bm I_b\bm\omega_b$
    \State Advance body-frame angular velocity:
    \[
      \bm\omega_b \gets
      \bm\omega_b + \Delta t\,\bm I_b^{-1}
      \left(\bm\tau_b-\bm\omega_b\times\bm L_b\right)
    \]
    \State Map updated angular velocity back to the world frame
  \Else
    \State Advance angular velocity with spherical inertia:
    \[
      \bm\omega \gets \bm\omega + \Delta t\,\bm\tau/(0.4mr^2)
    \]
  \EndIf

  \If{quaternion orientation is stored}
    \State Advance quaternion using world-frame angular velocity
    \State Normalize $\bm q$
  \EndIf
\EndFor
\end{algorithmic}
\end{algorithm}

\subsection{Simulations setup}
\label{sec:setup}
Three-dimensional simulations were performed in the context of an idealized cylindrical micro-vessel of diameter $d = 100, 150$ and  $200$ $\mu m$ and length $l=600$ $\mu m$, as depicted in Figure \ref{fig:setup}. Inlet and outlet boundaries were treated as periodic to reduce the computational domain and gain computational speed. Plasma flow was driven by a pressure drop $\Delta P$ over the channel length which ensured a particle Re within Stokesian regime. The variation of $\Delta P$ allowed an exploration of shear rate $\dot{\gamma}$, ranging from 150 to 1650 $s^{-1}$.

Here $\dot{\gamma}$ is computed as $d\Delta P/4\mu l$, with the wall shear stress $\tau_w=d\Delta P/4l$ fixed exactly by the streamwise momentum balance and $\mu$ the plasma viscosity. Since the vessel wall is bathed in the cell free layer, the near wall fluid is essentially pure plasma, thus this expression estimates the true wall shear rate rather than a nominal one. This definition is adopted consistently throughout. On the other hand, the apparent wall shear rate built from the realized mean velocity ($8\bar U/d$) would be smaller by exactly the suspension's relative apparent viscosity, $\dot\gamma/(8\bar U/d)=\mu_{app}/\mu\approx 1.4$ at $Ht=15\%$, consistent with the F\aa hr\oe us–Lindqvist reduction expected in a $100\,\mu m $ vessel \cite{Fahreus1931}, while the pseudo shear rate $\bar U/d$ adopted by some authors  would lie a further factor of eight below. These definitional factors must be reconciled throughout existing literature before threshold shear rates are compared.

For simplicity, every PLT was initially positioned randomly (Figure \ref{fig:setup}.b) following a uniform distribution along the channel length and oriented such that $\bm{q}=(1,0,0,0)$. As per Section \ref{subsec: orientation} the principal moments of inertia were $I_{xx}=I_{yy}=8.2\,pg\mu m^2$ and $I_{zz}= 15.4\,pg\mu m^2$. The number density of PLTs was such to satisfy the physiological average value of $300{,}000$ cells per $\mu l$ of blood, therefore it was considered independent from the number of RBCs in the simulation, which instead was adapted to satisfy an Ht\% of 10, 15 and 20\%.

Finally, it is worth noting that the geometry most frequently employed in the present work ($d = 100\,\mu m$ and $l=600\,\mu m$) required a total of only $1{,}536$ fluid cells when discretized with the \textit{unresolved} framework. In contrast, simulating the same geometry with a fully \textit{resolved} approach, requiring at least 8 cells per PLT diameter to ensure adequate resolution, would call for approximately 90 million fluid cells. This corresponds to a computational gain on the order of $\sim5.8\times10^4$, cementing the substantial efficiency of \textit{unresolved} CFD--DEM for large-scale domains.

\section{Results and Discussion}
The following paragraphs will focus on reporting how radial particle distribution, particle diffusion, PLTs margination, and CFL formation were impacted by channel geometry and flow conditions, via the exploration of a range of channel diameters and $\dot{\gamma}$. Different values of Ht\% were also investigated up to the 20\% range, due to limitations on the numerical consistency of the RBCs drag and lift models. In this regard, it should also be noted that hematocrit in the micro circulation is bound between 5 and 20\%, therefore it is not physiologically inaccurate to operate within this limit for tubular channels of $100\mu m$ in diameter \cite{Boyle1988}.

\subsection{Oblate force model}
\label{sec:oblates_results}
\begin{figure}
	\centering
	\includegraphics[width=1\columnwidth]{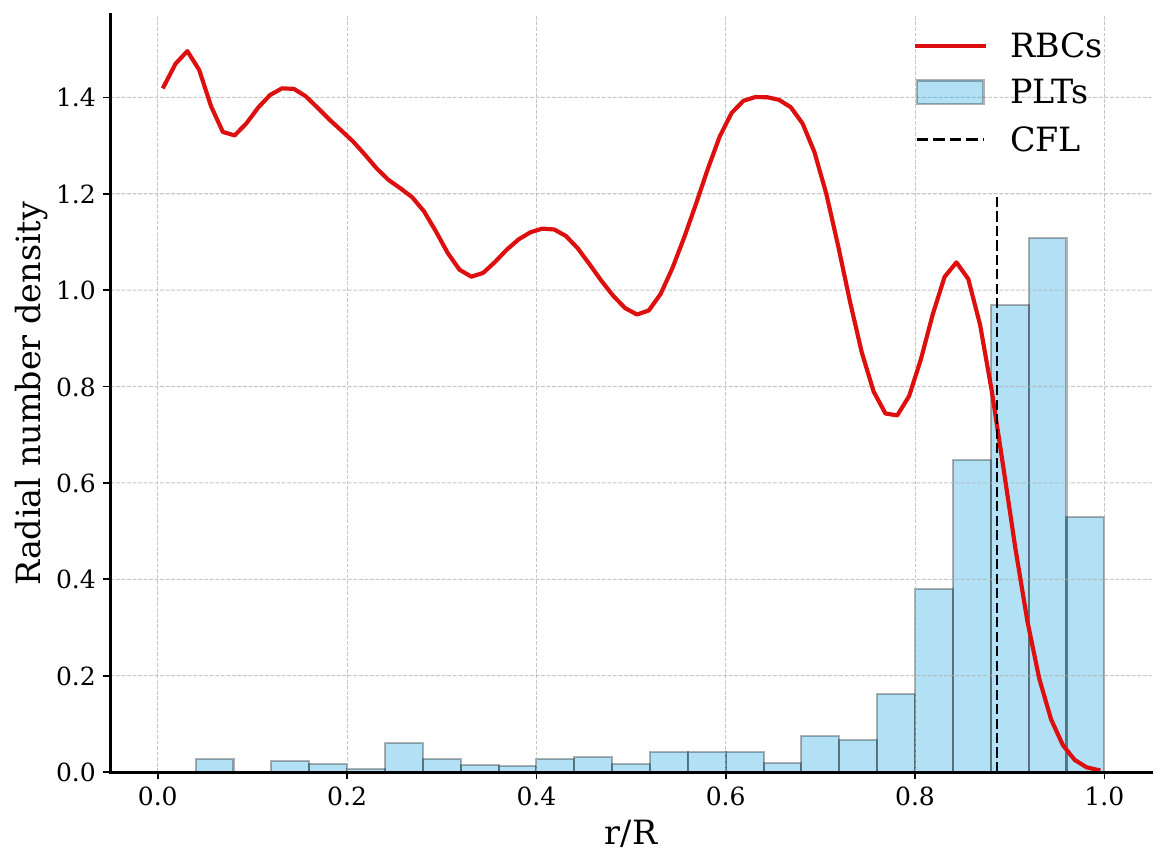}
	\caption{PLTs and RBCs radial number density profiles at time $t = 2s$ for $d = 100\,\mu m$,  $\dot{\gamma}=799\,s^{-1}$ and $Ht = 15\%$. The dashed line marks the CFL as per the definition given in Section \ref{sec:oblates_results}}
	\label{fig:oblate_rbc_distribution}
\end{figure}

The concentration profile of oblate PLTs and RBCs across the channel normalized radius is shown in Figure \ref{fig:oblate_rbc_distribution} at $t=2s$ for a $15\%Ht$ suspension containing approximately 9'400 particles. At this point in time, RBCs have migrated away from the wall and gathered in the core region, whereas platelets have accumulated in the near-wall cell free layer. The dashed line marks the CFL at $r/R\approx 0.88$ following the method of Ref. \cite{Yin2013}, as the community currently lacks a universally recognized way to define this region. The almost complete particle segregation achieved with this simulation confirms the ability of the current framework to satisfactorily reproduce the outcomes observed experimentally and in fully resolved simulations. Within the simplified world of this unresolved framework, the orientation dependent model in use for the drag/lift/torque of oblate particles in combination with RBC-induced hydrodynamic and collisional transport, is proven sufficient to capture the wall directed platelet drift.
The following sections will focus on a more quantitative validation of the framework.

\subsection{Effect of particle geometry}
\label{sec:oblateVSsphere}
\begin{figure*}
	\centering
	\includegraphics[width=1.\textwidth]{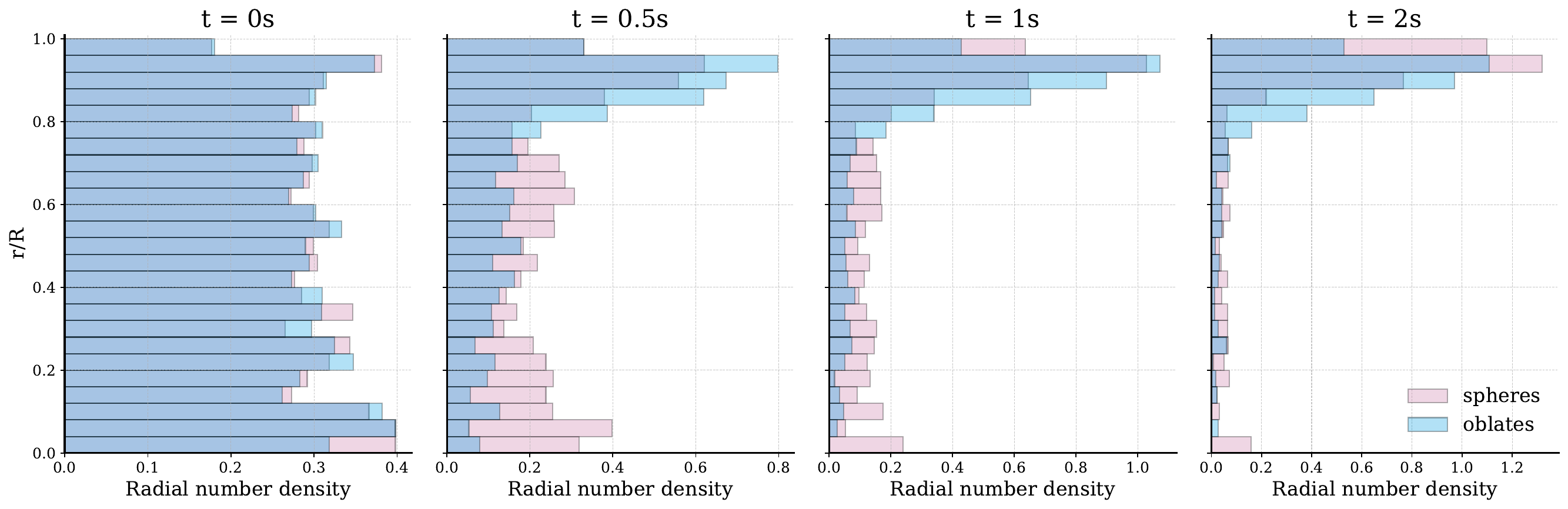}
	\caption{Comparison of platelet radial number density profiles at time $t = 0, 0.5, 1$ and $2s$ for $d = 100\,\mu m$, $\dot{\gamma}=799\,s^{-1}$ and $Ht = 15\%$. Profiles corresponding to oblate force models are shown in light blue, while those corresponding to spherical force models are shown in pink. Darker blue indicates the presence of both particle type. Note the changing range of the x axis.}
	\label{fig:oblatesVSspheres_distribution}
\end{figure*}
\begin{figure}
	\centering
	\includegraphics[width=1\columnwidth]{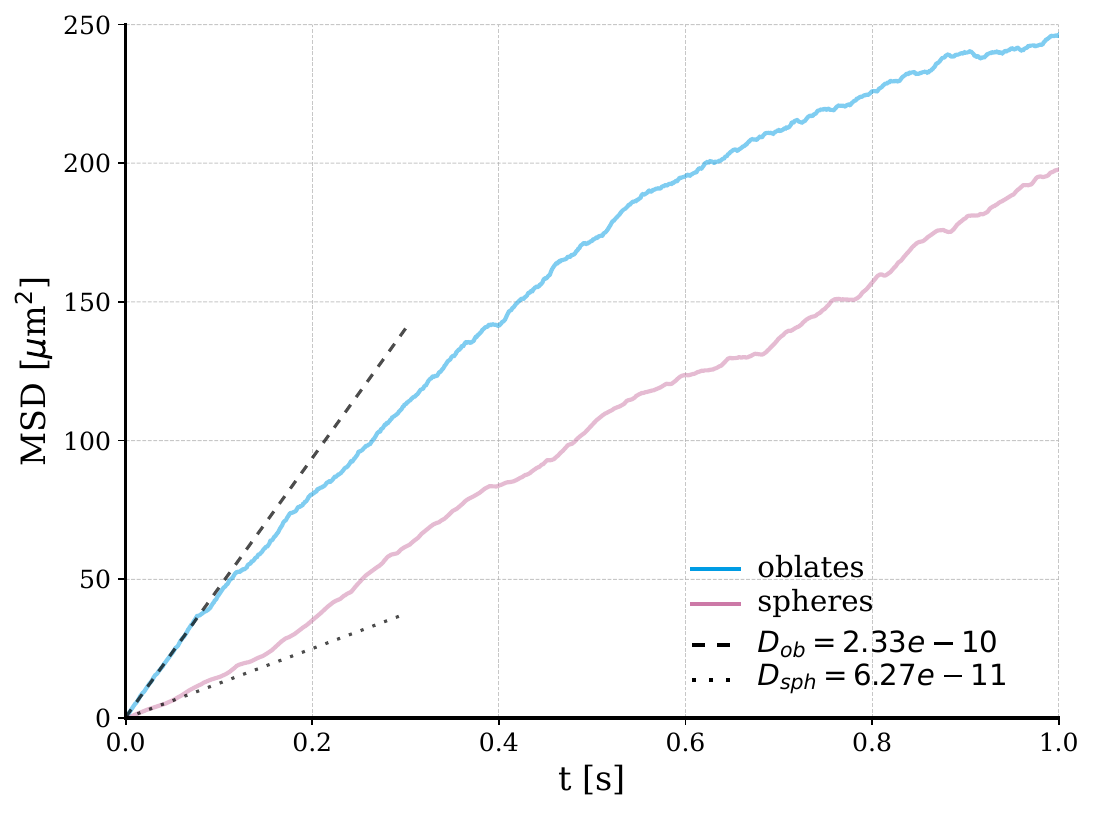}
	\caption{MSDs and corresponding diffusion coefficients of particles obeying the oblate force models (light blue) and the spherical force models (pink). Channel $d = 100\,\mu m$,  $\dot{\gamma}=799\,s^{-1}$ and $Ht = 15\%$. }
	\label{fig:msd_shape_comparison}
\end{figure}
Two separate simulations were set up with the same initial conditions to compare PLTs force models: the oblate dynamic (equations \ref{eq:CD_plt}, \ref{eq:CL_plt} and \ref{eq:CT_plt}) and the spherical dynamic (equations \ref{eq:CD_shiller} and \ref{eq:CL_shearlift}). The chosen geometry was again a cylindrical micro vessel of diameter $d=100\,\mu m$ with imposed wall shear rate of $\dot{\gamma}=799\,s^{-1}$. The $15\%Ht$ suspension evolved for a total of $2s$ of physical time while platelets were tracked to quantify margination. Figure \ref{fig:oblatesVSspheres_distribution} reports the radial density distribution of both types of particle models over time, well past the steady state point, when both species of particles have fully marginated. Most of the action happens within the first second of simulation, indeed the first three histograms of Figure \ref{fig:oblatesVSspheres_distribution} capture how differently the two force models evolve right from the start. Both simulations begin with particles uniformly distributed, at $t=0.5s$ the oblate model has already produced a more pronounced accumulation of particles near the channel wall with respect to the spherical counterpart, which is slowly catching up. The trend continues and at $t=1s$ it is evident that the oblate force models allow for a faster margination dynamic, as the amount of spherical particles still entrapped in the core is non negligible. Interestingly, the distribution at $t=2s$ shows that the spherical model finally catches up: within the bulk the two profiles are comparable while, closer to the CFL, the two distributions differ solely on the thickness of annular space claimed by the two species of PLTs. Specifically, CFL thickness is recorded at $5.7\,\mu m$ for the oblate model against a lower $3.9\, \mu m$ for the spherical one. This behavior suggests that the geometry of the particles and therefore the two forcing models, influence the speed of evolution of the segregation more than they does the final outcome, and clearly affect RBC central migration just as well. To further corroborate this hypothesis, the mean square displacement (MSD) for both simulations was computed as per the following:
\begin{align}
    MSD(t) = \langle [r_i(t) - r_i(0)]^2 \rangle_i ,
\end{align}
(where the $\langle\,\rangle_i$ mean an avarage over all PLTs \textit{i})
allowing for the derivation of PLTs diffusion coefficient, which is $D_{MSD}=MSD(t)/2t$ in a one dimensional system. The oblate model reported a diffusivity of $2.33 \times 10^{-10} \,m^2/s$ while the spherical model of $6.27 \times 10^{-11} \,m^2/s$, revealing a difference of almost one order of magnitude (Figure \ref{fig:msd_shape_comparison}). It appears that the spherical model underestimated particle diffusivity, slowing down the margination process, as it could be inferred by the distributions of Figure \ref{fig:oblatesVSspheres_distribution}. This finding is well in line with the work of Vahidkhah and Bagachi \cite{Vahidkhah2015} and by extension Zhao et al. \cite{Zhao2012}, which focused on blood flow simulations of spherical and oblate particles of different aspect ratios, showing how both species reach nearly the same radial distribution at steady state, despite proving the margination rate of particles to be geometry-specific.

To deepen this comparison, it is worth paying more careful attention to the diffusion coefficient. It is the natural metric governing the rate at which PLTs disperse towards the wall, hence the quantitative descriptor of margination. The order of magnitude obtained for the oblate model $O(10^{-10}\,m^2/s)$, is consistent with multiple shape resolved whole blood simulations \cite{Vahidkhah2015, Zhao2012}, which report PLT diffusivity in the same range and likewise find the margination rate to be geometry dependent while the steady state distribution is not, as discussed above. The spherical surrogate falls almost an order of magnitude below, mirroring the well documented tendency of continuum, shape-agnostic closures, such as the classical Zydney-Colton shear induced diffusivity \cite{Zydney1988}, to underestimate platelet transport by as much as two orders of magnitude at vessel scales once they are confronted with high fidelity cellular flow \cite{Kotsalos2022}.

The presented findings do not intend to invalidate the widespread use of spherical tracers to represent platelets, especially in experimental settings, but rather locate its limitations: a spherical description may recover the correct destination, though following a slower journey towards it. Moreover, the consequences of this limitation extend beyond PLTs alone, affecting also the red cell population. Being PLT and RBCs dynamically coupled, the slower and less anisotropic PLT drift, produced by the spherical model, feeds back onto the RBCs resulting in a weaker compaction of red cells towards the channel core with respect to the oblate counterpart. This in turn reshapes the CFL, changing its thickness, as stated earlier in the paragraph. The choice of platelet force model therefore influences not only the time scale of PLT margination but also, indirectly, the behaviour of RBCs.   

The next Sections will uniquely focus on PLTs following the oblate dynamic, intending to investigate the effects of channel size, shear rate and Ht\% on margination.

\subsection{Effect of channel diameter and shear rate}
\label{sec:wsr_effects}
\begin{figure}
	\centering
	\includegraphics[width=1\columnwidth]{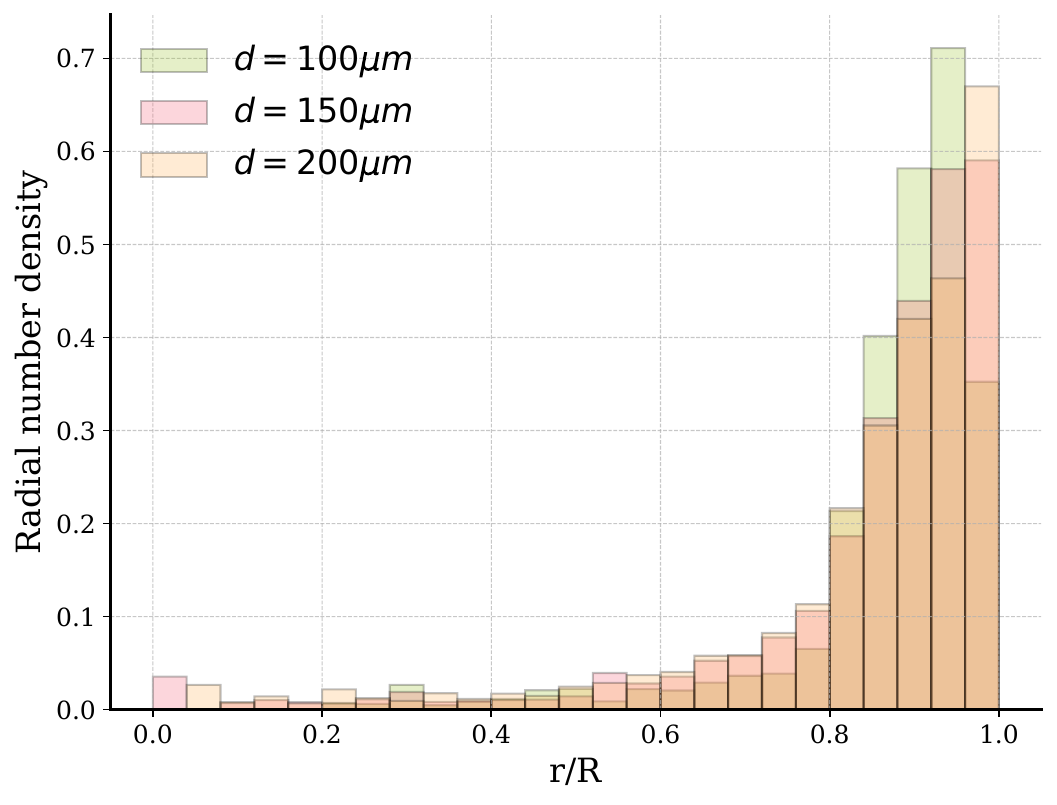}
	\caption{Comparison of platelet radial number density profiles of particles obeying the oblate force models at time $t = 2\,s$ for $d = 100, 150$ and $200\, \mu m$, $\dot{\gamma}=799\,s^{-1}$ and $Ht = 15\%$.}
	\label{fig:distribution_diam_comparison}
\end{figure}

\begin{figure}
	\centering
	\includegraphics[width=1\columnwidth]{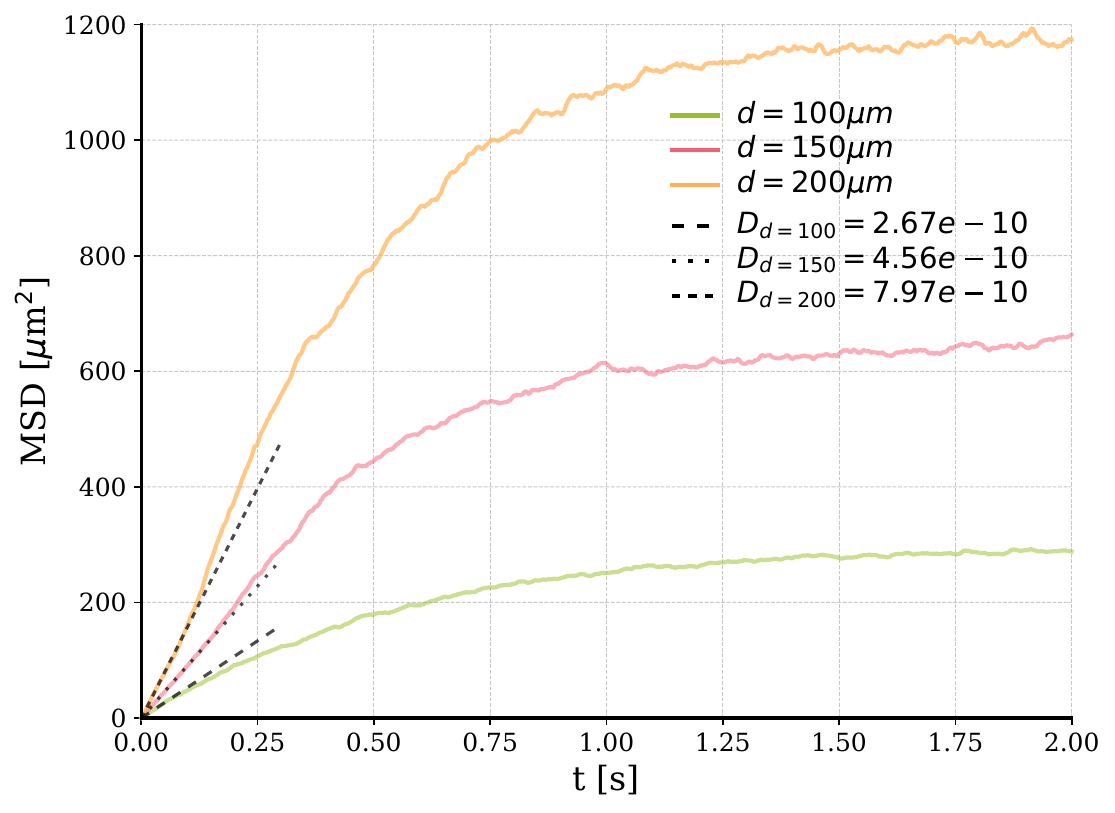}
	\caption{MSDs and corresponding diffusion coefficients comparison for particles obeying the oblate force model. Channel $d = 100, 150$ and $200\, \mu m$, $\dot{\gamma}=799\,s^{-1}$ and $Ht = 15\%$. }
	\label{fig:msd_diamComparison}
\end{figure}

First, three separate simulations were run imposing the same wall shear rate of $799\,s^{-1}$ on channels of diameter $d = 100, 150$ and  $200\,\mu m$ with $Ht = 15\%$. As it is well known, $\dot{\gamma}$ is one of the main driving parameters of margination, therefore it should follow that fixing its value would lead to similar PLT near-wall excesses despite the changing channel dimensions. The obtained results are reported in Figure \ref{fig:distribution_diam_comparison}, where the three PLT radial distributions have been normalized by the respective channel radius to aid the comparison. The histogram shows a very good match in terms of particles accumulated within the CFL. This result indicates that because of the identical prescribed wall shear rate, particles from all three simulations locally experience the same hydrodynamic landscape and therefore produce closely comparable normalized radial distributions. This behavior is achieved through an increased radial drift velocity in larger channels, as demonstrated in Figure \ref{fig:msd_diamComparison}. A less obvious yet noteworthy consequence concerns the CFL itself: the absolute thickness of the depleted annulus displays a dependence on the size of the system, thinning from $5.7\,\mu m$ in the $100\,\mu m$ channel to $5.1\,\mu m$ and $4.7\,\mu m$ as the diameter widens to $150\,\mu m$ and $200\,\mu m$. The near wall platelet excess is thus dictated by the local shear rate , whereas the absolute thickness of the CFL still reflects the overall dimensions of the vessel. Moreover, Figure  \ref{fig:msd_diamComparison} highlights the effect of flow confinement, indeed all three MSD curves flatten after some time, indicating that no other PLT would displace further. Mind that this flattening occurs at higher MSDs as the channel diameter increases, highlighting how diffusion closely depends on system dimensions. At the relatively high shear rates considered here, PLTs migrate rapidly toward the wall, resulting in the steep initial increase of the mean square displacement curves from which the corresponding diffusion coefficients are derived. After approximately $0.5s$, the MSD reaches a plateau, indicating that the population of particles capable of entering the CFL has essentially reached equilibrium, with only minor fluctuations around the steady state thereafter.

\begin{figure}
	\centering
	\includegraphics[width=1\columnwidth]{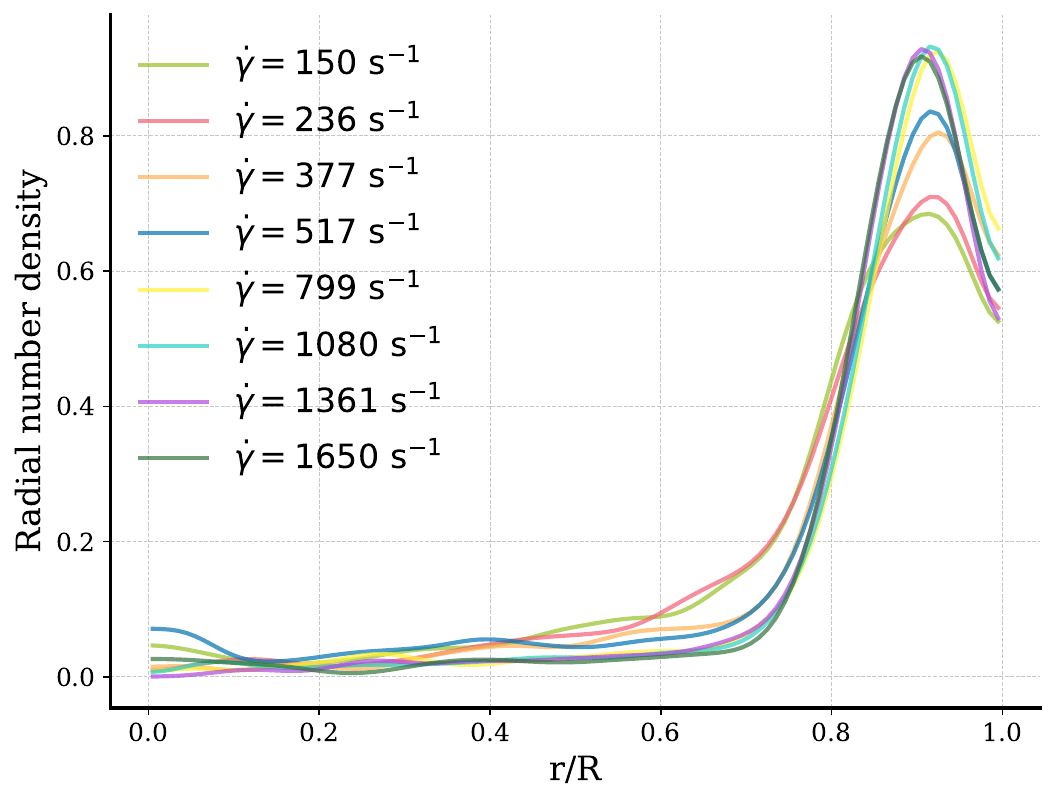}
	\caption{Comparison at steady state of platelet radial number density profiles of particles obeying the oblate force model for $d = 100\,\mu m$, $\dot{\gamma}= 150 - 1650\,s^{-1}$ and $Ht = 15\%$.}
	\label{fig:distribution_shear_comparison}
\end{figure}

\begin{figure}
	\centering
	\includegraphics[width=1\columnwidth]{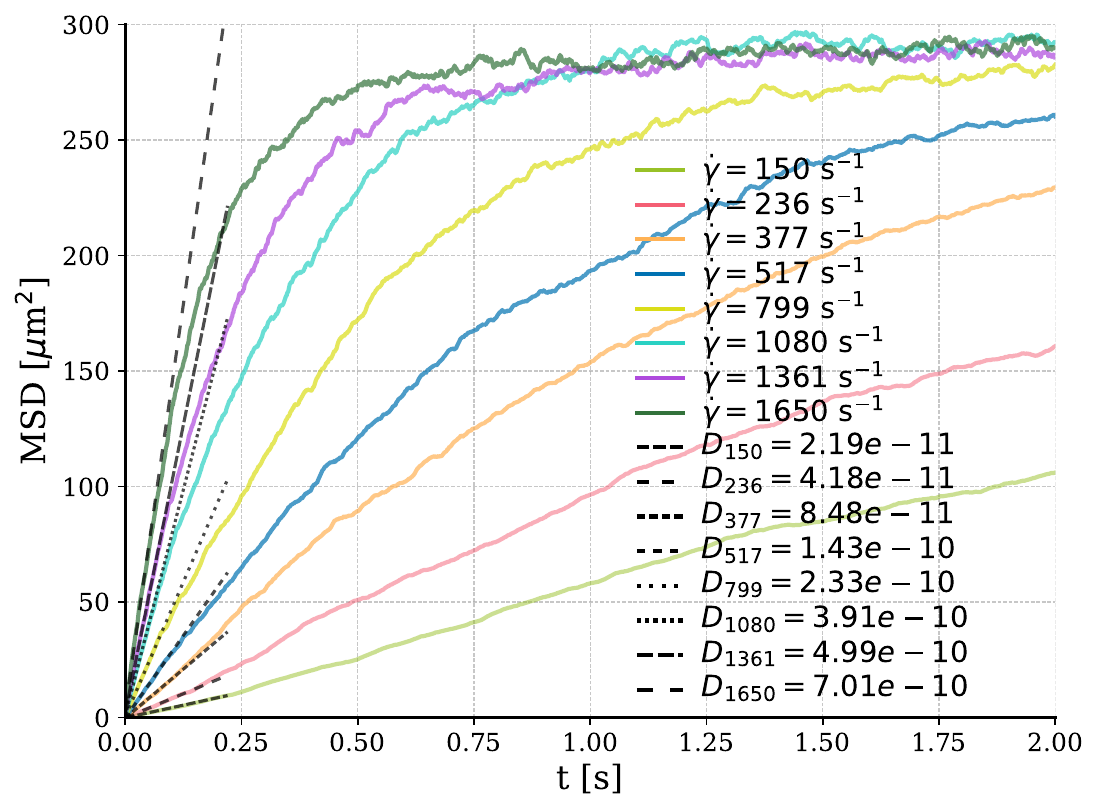}
	\caption{MSDs and corresponding diffusion coefficients of particles obeying the oblate force models. Channel $d = 100\,\mu m$,  $\dot{\gamma}=150 - 1650\,s^{-1}$ and $Ht = 15\%$. }
	\label{fig:msd_shear_comparison}
\end{figure}

\begin{table}[width=.9\linewidth,cols=3,pos=h]
\caption{CFL thickness across WSR values in a $100 \mu m$ channel at $Ht = 15\%$.}\label{tbl1}
\begin{tabular*}{\tblwidth}{@{} LLLL@{} }
\toprule
$\dot{\gamma}$ $[s^{-1}]$ & $8\bar U/d$ $[s^{-1}]$ & CFL $[\mu m]$\\
\midrule
150 & 135 & 3.1 \\
236 & 201 & 3.5 \\
377 & 285 & 4.4 \\
517 & 407 & 5.1 \\
799 & 570 & 5.7 \\
1080 & 804 & 6.5 \\
1361 & 940 & 7.6 \\
1650 & 1178 & 8.8 \\
\bottomrule
\end{tabular*}
\label{table:cfl_comparison}
\end{table}
Subsequently, eight separate simulations were run to explore $\dot{\gamma}$ values in the range $150-1650\,s^{-1}$, in a $100\,\mu m$ cylinder and for $Ht = 15\%$. Figure \ref{fig:distribution_shear_comparison} shows the radial density distribution at steady state of PLTs for increasing values of wall shear rate, whereas Figure \ref{fig:msd_shear_comparison} reports the corresponding evolution of MSD for the first $2s$ of all eight simulations, accompanied with diffusion coefficient data (not visible in Figure \ref{fig:msd_shear_comparison}: simulations at lower $\dot{\gamma}$ were carried out longer than $2s$, until steady state). It is possible to observe a clear separation of trends which groups the results in three categories. The cases at lower $\dot{\gamma}$ clearly show a wider spread of PLTs, which hold a thicker portion of annular vessel space partly shared with RBCs as the CFL (Table \ref{table:cfl_comparison}) is still quite thin in this regime. As $\dot{\gamma}$ increases to $799\,s^{-1}$ the number of particles close to the vessel wall increases as the RBC-PLT interactions become more frequent, a direct effect of higher shear rate, and PLTs are squeezed out through volume exclusion as RBCs compact towards the centerline, causing the CFL to thicken. Finally, the three simulations at $\dot{\gamma}=1080\,s^{-1}$ and above corresponding to $(8U/d)\geq804\,s^{-1}$, all settle around the same PLT density profile, independently of the applied shear rate. This indicates that the optimal shear rate maximizing margination has been reached \cite{Kruger2016, Li2023}, consistent with the findings of Freund and Orescain \cite{Freund2011}. Here, platelets reach their maximal crowding for the given Ht$\%$: despite the continuous increase in diffusion coefficient (Figure \ref{fig:msd_shear_comparison}), the density profiles in Figure \ref{fig:distribution_shear_comparison} converge onto one another corroborating the idea that this near-wall region cannot accommodate a higher particle density, even as particles reach the emptying space near the wall more quickly.
Given that the flux of marginating particles loses its dependency on shear rate above this $\dot{\gamma}=1080\,s^{-1}$ threshold, one might expect a concurrent arrest in the growing CFL thickness, since the two phenomena are interconnected and expected to follow a similar behavior. However, the data in Table \ref{table:cfl_comparison} does not support this expectation, and the most likely explanation lies in the RBC models used here. The controlling parameter for CFL thickness should not be the shear rate itself, but the RBC capillary number \cite{Kruger2016}: the CFL ceases to evolve once red cells reach their maximal deformation and no longer stretch with increasing shear, so that different reported shear-rate thresholds in fact correspond to a common capillary number. The present RBC drag and lift models lack this deformation dependency, which is believed to be responsible for the absence of an analogous threshold in CFL thickness in our simulations, making the presented CFL estimations reliable only within $\dot{\gamma}=799\,s^{-1}$. This threshold is quoted at different nominal values across the literature, owing precisely to the definitional ambiguity note in Section \ref{sec:setup}: Freund and Orescain \cite{Freund2011} place it near $\bar U/d \approx 100\,s^{-1}$, equivalent to $8\bar U/d \approx 800\,s^{-1}$, whereas Ye et al. \cite{Ye2018} report $\approx 1000\,s^{-1}$. Expressed on a common wall-shear basis, the two values are mutually consistent and both bracket the regime change observed here, supporting the interpretation that the apparent disagreement in the literature is one of definition rather than of underlying physics.

\subsection{Effect of hematocrit }
\begin{figure}
	\centering
	\includegraphics[width=.8\columnwidth]{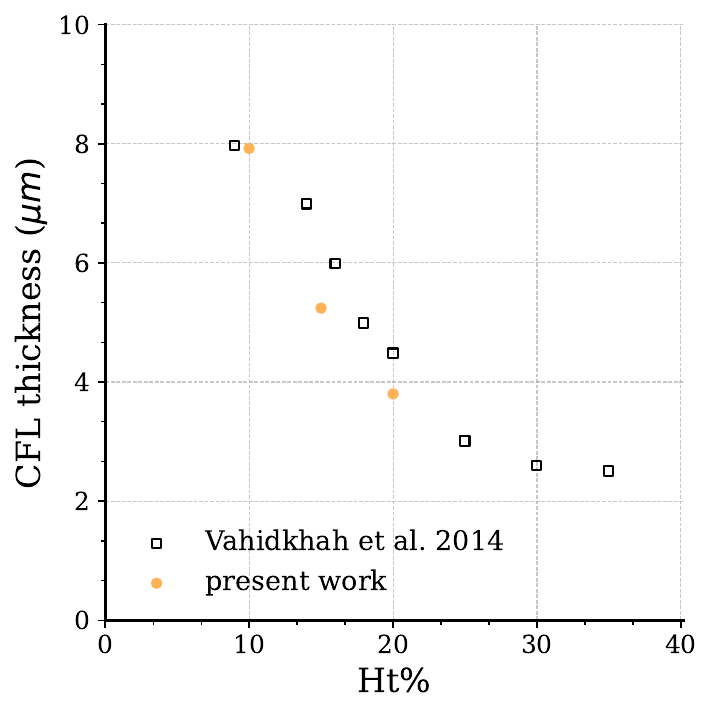}
	\caption{Thickness of RBC depleted region (CFL) as a function of Ht\%. Values from the present work shown as orange dots, comparison with the work of Vahidkhah et al. \cite{Vahidkhah2014} as black squares.}
	\label{fig:cfl_ht_comparison}
\end{figure}
The last parameter explored in this work is hematocrit. Three suspensions with $Ht = 10, 15$ and $20\%$ were evolved in a $100\,\mu m$ cylinder at a fixed wall shear rate of $799\,s^{-1}$, so that any change in PLT distribution could be ascribed to the red cell volume fraction alone. The thickness of the RBC depleted region was extracted at steady state  and is reported in Figure \ref{fig:cfl_ht_comparison} as a function of $Ht\%$. The CFL thins monotonically as hematocrit grows, contracting from roughly $7.9$ to $3.8\,\mu m$ almost halving as Ht rises from $10$ to $20\%$. This is the signature of a well documented phenomenon: as the core of a vessel becomes more crowded, the volume exclusion effect that RBCs exert on PLTs intensifies, packing red cells into a tighter central column while driving a larger fraction of PLTs outward \cite{TillesEckstein1987, Reasor2013, Zhao2012}. The same crowding that thins the depleted annulus is what sharpens the platelet peak that sits within it, so that a higher $Ht\%$ acts on both species at once. This dependence is consistent with the historical (recorded) picture of margination as an inherently RBC-driven process, observed to vanish below a few percent hematocrit and to strengthen steeply with Ht in the early channel experiments of Tilles and Eckstein \cite{TillesEckstein1987, Eckstein1988} and absent altogether in RBC free suspensions \cite{Turitto1975, Aarts1988}. The obtained results closely follow the whole blood simulations of Vahidkhah et al. \cite{Vahidkhah2014}, who report the CFL thinning from approximately $8$ to $2.5\,\mu m$ as hematocrit is raised from $9$ to $36\%$. The two datasets coincide at the lowest Ht while loosely departing over the $15-20\%$ range, with their CFL values sitting marginally above. This small offset is attributed to the difference in imposed WSR. Their suspensions were driven at $1000\,s^{-1}$ and as cited at the end of Section \ref{sec:wsr_effects}, CFL thickness is sensitive to shear rate up until the $\dot{\gamma}=1000\,s^{-1}$ threshold, above which RBCs cease to deform further. At the current $\dot{\gamma}=799\,s^{-1}$ the layer is therefore still slightly thinner than its theoretical saturated value. Within the narrow hematocrit window accessible to the present RBC drag and lift closures, the model thus recovers the expected relation between Ht and CFL thickness, closing on its ability to reproduce the joint dependence of PLT margination on shear rate and red cell concentration.

By contrast, platelet diffusivity extracted from the MSD showed no systematic dependence on hematocrit over the explored window, remaining of the order $2.5\times10^{-10}\,m^2/s$ ($2.75$, $2.33$ and $2.54\times10^{-10} \, m^2/s$ at $Ht=10$, $15$ and $20\%$). This clearly contradics the prediction of the Zydney-Colton's law, as the increase in diffusion coefficient predicted by the law is not observed over the explored Ht\% range \cite{Zydney1988}. At a fixed WSR and channel size, and within this low physiological range, it is the red cell fraction which dictates the geometry of the depleted layer far more directly than the diffusive transport of PLTs, whose magnitude is here governed primarily by the shear rate, as see in Section \ref{sec:wsr_effects}. The collision-driven rise of PLTs diffusivity with hematocrit reported in fully resolved whole blood simulations \cite{Vahidkhah2014, CrowlFogelson2011} develops over considerably wider Ht ranges and is not resolvable within the $10-20\%$ window accessible to the present closures.

\section{Conclusion}
This work presented an unresolved DEM model for platelet dynamics and integrated it into an unresolved CFD-DEM solver for blood flow simulation based on OpenFOAM-LIGGGHTS coupling, completing the cellular description of an existing unresolved RBC framework \cite{PorcaroSaeedipour2024}. Within this mesoscale paradigm, platelets were represented as rigid oblate particles equipped with quaternion based rotational dynamics and the corresponding anisotropic inertia tensor, while their hydrodynamic response was supplied by orientation dependent drag, lift and torque closures derived from oblate ellipsoids. Despite the strong simplification inherent to the unresolved approach, most notably the local uniform flow approximation of the hydrodynamic forces, the model reproduced the salient features of platelet transport in a cylindrical microvessel populated with RBCs: the spontaneous segregation of the two species into an RBC-rich core and a PLT-rich cell free layer, the geometry specific margination dynamics of oblate versus spherical particles, the control exerted by wall shear rate together with its saturation above the deformability threshold, and the inverse dependence of cell free layer thickness on hematocrit, in close agreement with resolved whole blood simulations.

Taken together, these simulations do more than confirm known phenomenology; they allow two quantities of direct hemorheological interest, the PLT diffusion coefficient and CFL thickness, to be tracked across the whole computational window and their parametric dependencies to be read off explicitly. The platelet diffusivity, of order $10^{-10}\,m/s$ throughout, was found to grow both with the size of the system and with the imposed wall shear rate, while remaining essentially insensitive to Ht over the 10--20\% range explored, a notable departure from the increase predicted by shear-induced closures of the Zydney–Colton type. The CFL emerged as a function of the dimensions of the system, thickening with wall shear rate and thinning monotonically with Ht\% as the crowded red cell core drives a growing fraction of PLTs outward. That both quantities could be resolved with this level of parametric detail and across a combination of vessel sizes and shear range seldom explored together, is a direct consequence of the unresolved formulation. Indeed, what makes such a broad exploration affordable is the very coarsening that sacrifices fully resolved fidelity, and the dependencies recovered here remain meaningful within the stated limitations of the underlying RBC drag and lift closures.

Beyond this, the comparison between the oblate and the spherical model in Section \ref{sec:oblateVSsphere} carries a broader message. The two descriptions converged toward almost identical steady state distributions, yet they traveled towards equilibrium along very different trajectories, the oblate particle diffusing nearly and order of magnitude faster and marginating correspondingly sooner. The final outcome of the segregation process therefore seems to be forgiving of the modeling choice, but its temporal evolution is not. This distinction is of practical consequence since platelet margination, together with the hemostatic response it triggers, are inherently kinetic phenomena for which the relevant metric is not only where platelets end up but how quickly they get there. When the time evolution of a process holds as much importance as its endpoint, the predictive value of a simulation depends on how faithfully the particle force model represents the underlying physics. A closure that accounts for the anisotropy, orientation and torque driven rotation of the real cell can reproduce the transient dynamics observed physiologically, whereas a spherical approximation may recover the correct steady state, while misrepresenting the rate at which it is reached.

The present framework is not without limitations, and these naturally trace the direction of future work. The local closures neglect shear induced and many body hydrodynamic interactions as well as confinement corrections. The accessible hematocrit range remains bound by the current RBC drag and lift models which also lack deformation dependency; extending their validity would open the door to higher physiological and pathological hematocrits alike, as incorporating the capillary number would make CFL estimations reliable even in high shear regimes. Platelets are moreover treated as rigid bodies in their discoid shape, such that activation, shape change and adhesion are left to subsequent developments. Coupling the current model with activation and aggregation mechanics, and deploying it in anatomically realistic geometries, would turn the scalable vessel scale capability demonstrated here into a predictive tool for the study of thrombus initiation and growth. 

\bibliographystyle{model1-num-names}

\bibliography{cas-refs}

\begin{thebibliography}{46}
\expandafter\ifx\csname natexlab\endcsname\relax\def\natexlab#1{#1}\fi
\providecommand{\bibinfo}[2]{#2}
\ifx\xfnm\relax \def\xfnm[#1]{\unskip,\space#1}\fi
%Type = Book
\bibitem[{Walker et~al.(1990)Walker, Hall, and Hurst}]{Walker1990}
\bibinfo{editor}{H.~K. Walker}, \bibinfo{editor}{W.~D. Hall},
  \bibinfo{editor}{J.~W. Hurst} (Eds.), \bibinfo{title}{Clinical Methods: The
  History, Physical, and Laboratory Examinations. 3rd ed.},
  \bibinfo{publisher}{Boston: Butterworths}, \bibinfo{year}{1990}.
%Type = Inbook
\bibitem[{Thon and Italiano(2012)}]{Thon2012}
\bibinfo{author}{J.~N. Thon}, \bibinfo{author}{J.~E. Italiano},
  \bibinfo{title}{Platelets: Production, Morphology and Ultrastructure},
  \bibinfo{publisher}{Springer Berlin Heidelberg}, \bibinfo{address}{Berlin,
  Heidelberg}, pp. \bibinfo{pages}{3--22}.
%Type = Article
\bibitem[{Van~Hinsbergh(2012)}]{VanHinsbergh2012}
\bibinfo{author}{V.~W.~M. Van~Hinsbergh},
\newblock \bibinfo{title}{Endothelium---role in regulation of coagulation and
  inflammation},
\newblock \bibinfo{journal}{Seminars in Immunopathology} \bibinfo{volume}{34}
  (\bibinfo{year}{2012}) \bibinfo{pages}{93--106}.
%Type = Article
\bibitem[{Turitto and Baumgartner(1975)}]{Turitto1975}
\bibinfo{author}{V.~T. Turitto}, \bibinfo{author}{H.~R. Baumgartner},
\newblock \bibinfo{title}{Platelet interaction with subendothelium in a
  perfusion system: {{Physical}} role of red blood cells},
\newblock \bibinfo{journal}{Microvascular Research} \bibinfo{volume}{9}
  (\bibinfo{year}{1975}) \bibinfo{pages}{335--344}.
%Type = Book
\bibitem[{Turitto and Goldsmith(1996)}]{Turitto1996}
\bibinfo{author}{V.~T. Turitto}, \bibinfo{author}{H.~L. Goldsmith},
  \bibinfo{title}{Rheology, Transport and Thrombosis in the
  Circulation.{{Textbook}} of {{Cardiovascular Medicine}}},
  \bibinfo{publisher}{Lippincott Williams \& Wilkins}, \bibinfo{year}{1996}.
%Type = Article
\bibitem[{F{\aa}hr{\ae}us and Lindqvist(1931)}]{Fahreus1931}
\bibinfo{author}{R.~F{\aa}hr{\ae}us}, \bibinfo{author}{T.~Lindqvist},
\newblock \bibinfo{title}{{{THE VISCOSITY OF THE BLOOD IN NARROW CAPILLARY
  TUBES}}},
\newblock \bibinfo{journal}{American Journal of Physiology-Legacy Content}
  \bibinfo{volume}{96} (\bibinfo{year}{1931}) \bibinfo{pages}{562--568}.
%Type = Article
\bibitem[{Aarts et~al.(1988)Aarts, Van Den~Broek, Prins, Kuiken, Sixma, and
  Heethaar}]{Aarts1988}
\bibinfo{author}{P.~A. Aarts}, \bibinfo{author}{S.~A. Van Den~Broek},
  \bibinfo{author}{G.~W. Prins}, \bibinfo{author}{G.~D. Kuiken},
  \bibinfo{author}{J.~J. Sixma}, \bibinfo{author}{R.~M. Heethaar},
\newblock \bibinfo{title}{Blood platelets are concentrated near the wall and
  red blood cells, in the center in flowing blood.},
\newblock \bibinfo{journal}{Arteriosclerosis: An Official Journal of the
  American Heart Association, Inc.} \bibinfo{volume}{8} (\bibinfo{year}{1988})
  \bibinfo{pages}{819--824}.
%Type = Article
\bibitem[{Tilles and Eckstein(1987)}]{TillesEckstein1987}
\bibinfo{author}{A.~W. Tilles}, \bibinfo{author}{E.~C. Eckstein},
\newblock \bibinfo{title}{The near-wall excess of platelet-sized particles in
  blood flow: {{Its}} dependence on hematocrit and wall shear rate},
\newblock \bibinfo{journal}{Microvascular Research} \bibinfo{volume}{33}
  (\bibinfo{year}{1987}) \bibinfo{pages}{211--223}.
%Type = Article
\bibitem[{Reasor et~al.(2013)Reasor, Mehrabadi, Ku, and Aidun}]{Reasor2013}
\bibinfo{author}{D.~A. Reasor}, \bibinfo{author}{M.~Mehrabadi},
  \bibinfo{author}{D.~N. Ku}, \bibinfo{author}{C.~K. Aidun},
\newblock \bibinfo{title}{Determination of {{Critical Parameters}} in
  {{Platelet Margination}}},
\newblock \bibinfo{journal}{Annals of Biomedical Engineering}
  \bibinfo{volume}{41} (\bibinfo{year}{2013}) \bibinfo{pages}{238--249}.
%Type = Article
\bibitem[{Ye et~al.(2018)Ye, Shen, and Li}]{Ye2018}
\bibinfo{author}{H.~Ye}, \bibinfo{author}{Z.~Shen}, \bibinfo{author}{Y.~Li},
\newblock \bibinfo{title}{Shear rate dependent margination of sphere-like,
  oblate-like and prolate-like micro-particles within blood flow},
\newblock \bibinfo{journal}{Soft Matter} \bibinfo{volume}{14}
  (\bibinfo{year}{2018}) \bibinfo{pages}{7401--7419}.
%Type = Article
\bibitem[{Chang et~al.(2018)Chang, Yazdani, Li, Douglas, Mantzoros, and
  Karniadakis}]{Chang2018}
\bibinfo{author}{H.-Y. Chang}, \bibinfo{author}{A.~Yazdani},
  \bibinfo{author}{X.~Li}, \bibinfo{author}{K.~A. Douglas},
  \bibinfo{author}{C.~S. Mantzoros}, \bibinfo{author}{G.~E. Karniadakis},
\newblock \bibinfo{title}{Quantifying {{Platelet Margination}} in {{Diabetic
  Blood Flow}}},
\newblock \bibinfo{journal}{Biophysical Journal} \bibinfo{volume}{115}
  (\bibinfo{year}{2018}) \bibinfo{pages}{1371--1382}.
%Type = Article
\bibitem[{Eckstein et~al.(1988)Eckstein, Tilles, and Millero}]{Eckstein1988}
\bibinfo{author}{E.~C. Eckstein}, \bibinfo{author}{A.~W. Tilles},
  \bibinfo{author}{F.~J. Millero},
\newblock \bibinfo{title}{Conditions for the occurrence of large near-wall
  excesses of small particles during blood flow},
\newblock \bibinfo{journal}{Microvascular Research} \bibinfo{volume}{36}
  (\bibinfo{year}{1988}) \bibinfo{pages}{31--39}.
%Type = Article
\bibitem[{Kr{\"u}ger(2016)}]{Kruger2016}
\bibinfo{author}{T.~Kr{\"u}ger},
\newblock \bibinfo{title}{Effect of tube diameter and capillary number on
  platelet margination and near-wall dynamics},
\newblock \bibinfo{journal}{Rheologica Acta} \bibinfo{volume}{55}
  (\bibinfo{year}{2016}) \bibinfo{pages}{511--526}.
%Type = Article
\bibitem[{Li et~al.(2023)Li, Wang, Han, Qi, Ma, Li, Yin, Li, Li, and
  Qian}]{Li2023}
\bibinfo{author}{L.~Li}, \bibinfo{author}{S.~Wang}, \bibinfo{author}{K.~Han},
  \bibinfo{author}{X.~Qi}, \bibinfo{author}{S.~Ma}, \bibinfo{author}{L.~Li},
  \bibinfo{author}{J.~Yin}, \bibinfo{author}{D.~Li}, \bibinfo{author}{X.~Li},
  \bibinfo{author}{J.~Qian},
\newblock \bibinfo{title}{Quantifying {{Shear-induced Margination}} and
  {{Adhesion}} of {{Platelets}} in {{Microvascular Blood Flow}}},
\newblock \bibinfo{journal}{Journal of Molecular Biology} \bibinfo{volume}{435}
  (\bibinfo{year}{2023}) \bibinfo{pages}{167824}.
%Type = Article
\bibitem[{Freund and Orescanin(2011)}]{Freund2011}
\bibinfo{author}{J.~B. Freund}, \bibinfo{author}{M.~M. Orescanin},
\newblock \bibinfo{title}{Cellular flow in a small blood vessel},
\newblock \bibinfo{journal}{Journal of Fluid Mechanics} \bibinfo{volume}{671}
  (\bibinfo{year}{2011}) \bibinfo{pages}{466--490}.
%Type = Article
\bibitem[{Corattiyl and Eckstein(1986)}]{CorattiylEckstein1986}
\bibinfo{author}{V.~Corattiyl}, \bibinfo{author}{E.~C. Eckstein},
\newblock \bibinfo{title}{Regional platelet concentration in blood flow through
  capillary tubes},
\newblock \bibinfo{journal}{Microvascular Research} \bibinfo{volume}{32}
  (\bibinfo{year}{1986}) \bibinfo{pages}{261--270}.
%Type = Article
\bibitem[{Dynar et~al.(2024)Dynar, {Ez-Zahraouy}, Misbah, and
  Abbasi}]{Dynar2024}
\bibinfo{author}{M.~Dynar}, \bibinfo{author}{H.~{Ez-Zahraouy}},
  \bibinfo{author}{C.~Misbah}, \bibinfo{author}{M.~Abbasi},
\newblock \bibinfo{title}{Platelet margination dynamics in blood flow: {{The}}
  role of lift forces and red blood cells aggregation},
\newblock \bibinfo{journal}{Physical Review Fluids} \bibinfo{volume}{9}
  (\bibinfo{year}{2024}) \bibinfo{pages}{083603}.
%Type = Article
\bibitem[{Sorensen et~al.(1999{\natexlab{a}})Sorensen, Burgreen, Wagner, and
  Antaki}]{Sorensen1999a}
\bibinfo{author}{E.~N. Sorensen}, \bibinfo{author}{G.~W. Burgreen},
  \bibinfo{author}{W.~R. Wagner}, \bibinfo{author}{J.~F. Antaki},
\newblock \bibinfo{title}{Computational {{Simulation}} of {{Platelet
  Deposition}} and {{Activation}}: {{I}}. {{Model Development}} and
  {{Properties}}},
\newblock \bibinfo{journal}{Annals of Biomedical Engineering}
  \bibinfo{volume}{27} (\bibinfo{year}{1999}{\natexlab{a}})
  \bibinfo{pages}{436--448}.
%Type = Article
\bibitem[{Sorensen et~al.(1999{\natexlab{b}})Sorensen, Burgreen, Wagner, and
  Antaki}]{Sorensen1999b}
\bibinfo{author}{E.~N. Sorensen}, \bibinfo{author}{G.~W. Burgreen},
  \bibinfo{author}{W.~R. Wagner}, \bibinfo{author}{J.~F. Antaki},
\newblock \bibinfo{title}{Computational {{Simulation}} of {{Platelet
  Deposition}} and {{Activation}}: {{II}}. {{Results}} for {{Poiseuille Flow}}
  over {{Collagen}}},
\newblock \bibinfo{journal}{Annals of Biomedical Engineering}
  \bibinfo{volume}{27} (\bibinfo{year}{1999}{\natexlab{b}})
  \bibinfo{pages}{449--458}.
%Type = Article
\bibitem[{Cardillo and Barakat(2025)}]{Cardillo2025}
\bibinfo{author}{G.~Cardillo}, \bibinfo{author}{A.~I. Barakat},
\newblock \bibinfo{title}{A {{2D}} computational model of chemically- and
  mechanically-induced platelet plug formation},
\newblock \bibinfo{journal}{Biomechanics and Modeling in Mechanobiology}
  \bibinfo{volume}{24} (\bibinfo{year}{2025}) \bibinfo{pages}{1465--1484}.
%Type = Article
\bibitem[{Crowl and Fogelson(2010)}]{CrowlFogelson2010}
\bibinfo{author}{L.~M. Crowl}, \bibinfo{author}{A.~L. Fogelson},
\newblock \bibinfo{title}{Computational model of whole blood exhibiting lateral
  platelet motion induced by red blood cells},
\newblock \bibinfo{journal}{International Journal for Numerical Methods in
  Biomedical Engineering} \bibinfo{volume}{26} (\bibinfo{year}{2010})
  \bibinfo{pages}{471--487}.
%Type = Article
\bibitem[{Crowl and Fogelson(2011)}]{CrowlFogelson2011}
\bibinfo{author}{L.~Crowl}, \bibinfo{author}{A.~L. Fogelson},
\newblock \bibinfo{title}{Analysis of mechanisms for platelet near-wall excess
  under arterial blood flow conditions},
\newblock \bibinfo{journal}{Journal of Fluid Mechanics} \bibinfo{volume}{676}
  (\bibinfo{year}{2011}) \bibinfo{pages}{348--375}.
%Type = Article
\bibitem[{Kotsalos et~al.(2022)Kotsalos, Raynaud, L{\"a}tt, Dutta, Dubois,
  Zouaoui~Boudjeltia, and Chopard}]{Kotsalos2022}
\bibinfo{author}{C.~Kotsalos}, \bibinfo{author}{F.~Raynaud},
  \bibinfo{author}{J.~L{\"a}tt}, \bibinfo{author}{R.~Dutta},
  \bibinfo{author}{F.~Dubois}, \bibinfo{author}{K.~Zouaoui~Boudjeltia},
  \bibinfo{author}{B.~Chopard},
\newblock \bibinfo{title}{Shear induced diffusion of platelets revisited},
\newblock \bibinfo{journal}{Frontiers in Physiology} \bibinfo{volume}{13}
  (\bibinfo{year}{2022}) \bibinfo{pages}{985905}.
%Type = Article
\bibitem[{Yazdani and Karniadakis(2016)}]{Yazdani2016}
\bibinfo{author}{A.~Yazdani}, \bibinfo{author}{G.~E. Karniadakis},
\newblock \bibinfo{title}{Sub-cellular modeling of platelet transport in blood
  flow through microchannels with constriction},
\newblock \bibinfo{journal}{Soft Matter} \bibinfo{volume}{12}
  (\bibinfo{year}{2016}) \bibinfo{pages}{4339--4351}.
%Type = Article
\bibitem[{Fedosov et~al.(2010)Fedosov, Caswell, Popel, and
  Karniadakis}]{Fedosov2010}
\bibinfo{author}{D.~A. Fedosov}, \bibinfo{author}{B.~Caswell},
  \bibinfo{author}{A.~S. Popel}, \bibinfo{author}{G.~E. Karniadakis},
\newblock \bibinfo{title}{Blood {{Flow}} and {{Cell-Free Layer}} in
  {{Microvessels}}: {{Blood Flow}} and {{Cell-Free Layer}} in
  {{Microvessels}}},
\newblock \bibinfo{journal}{Microcirculation} \bibinfo{volume}{17}
  (\bibinfo{year}{2010}) \bibinfo{pages}{615--628}.
%Type = Article
\bibitem[{Z{\'a}vodszky et~al.(2019)Z{\'a}vodszky, van Rooij, Czaja, Azizi,
  de~Kanter, and Hoekstra}]{Zavodszky2019}
\bibinfo{author}{G.~Z{\'a}vodszky}, \bibinfo{author}{B.~van Rooij},
  \bibinfo{author}{B.~Czaja}, \bibinfo{author}{V.~Azizi},
  \bibinfo{author}{D.~de~Kanter}, \bibinfo{author}{A.~G. Hoekstra},
\newblock \bibinfo{title}{Red blood cell and platelet diffusivity and
  margination in the presence of cross-stream gradients in blood flows},
\newblock \bibinfo{journal}{Physics of Fluids} \bibinfo{volume}{31}
  (\bibinfo{year}{2019}) \bibinfo{pages}{031903}.
%Type = Article
\bibitem[{Zhang et~al.(2017)Zhang, Zhang, Slepian, Deng, and
  Bluestein}]{Zhang2017}
\bibinfo{author}{P.~Zhang}, \bibinfo{author}{L.~Zhang}, \bibinfo{author}{M.~J.
  Slepian}, \bibinfo{author}{Y.~Deng}, \bibinfo{author}{D.~Bluestein},
\newblock \bibinfo{title}{A multiscale biomechanical model of platelets:
  {{Correlating}} with in-vitro results},
\newblock \bibinfo{journal}{Journal of Biomechanics} \bibinfo{volume}{50}
  (\bibinfo{year}{2017}) \bibinfo{pages}{26--33}.
%Type = Article
\bibitem[{Kloss et~al.(2012)Kloss, Goniva, Hager, Amberger, and
  Pirker}]{Kloss2012}
\bibinfo{author}{C.~Kloss}, \bibinfo{author}{C.~Goniva},
  \bibinfo{author}{A.~Hager}, \bibinfo{author}{S.~Amberger},
  \bibinfo{author}{S.~Pirker},
\newblock \bibinfo{title}{Models, algorithms and validation for opensource
  {{DEM}} and {{CFD-DEM}}},
\newblock \bibinfo{journal}{Progress in Computational Fluid Dynamics, An
  International Journal} \bibinfo{volume}{12} (\bibinfo{year}{2012})
  \bibinfo{pages}{140}.
%Type = Article
\bibitem[{Kotsalos et~al.(2019)Kotsalos, Latt, and Chopard}]{Kotsalos2019}
\bibinfo{author}{C.~Kotsalos}, \bibinfo{author}{J.~Latt},
  \bibinfo{author}{B.~Chopard},
\newblock \bibinfo{title}{Bridging the computational gap between mesoscopic and
  continuum modeling of red blood cells for fully resolved blood flow},
\newblock \bibinfo{journal}{Journal of Computational Physics}
  \bibinfo{volume}{398} (\bibinfo{year}{2019}) \bibinfo{pages}{108905}.
%Type = Article
\bibitem[{Porcaro and Saeedipour(2024)}]{PorcaroSaeedipour2024}
\bibinfo{author}{C.~Porcaro}, \bibinfo{author}{M.~Saeedipour},
\newblock \bibinfo{title}{Unresolved {{RBCs}}: {{An}} upscaling strategy for
  the {{CFD-DEM}} simulation of blood flow with deformable cells},
\newblock \bibinfo{journal}{Computers in Biology and Medicine}
  \bibinfo{volume}{181} (\bibinfo{year}{2024}) \bibinfo{pages}{109081}.
%Type = Article
\bibitem[{Porcaro et~al.(2025)Porcaro, Latt, and
  Saeedipour}]{PorcaroSaeedipour2025}
\bibinfo{author}{C.~Porcaro}, \bibinfo{author}{J.~Latt},
  \bibinfo{author}{M.~Saeedipour},
\newblock \bibinfo{title}{Bridging the scales for red blood cells simulation:
  From immersed boundary {{LBM-npFEM}} to unresolved {{CFD-DEM}}}
  (\bibinfo{year}{2025}).
%Type = Article
\bibitem[{Ouchene(2020)}]{Ouchene2020}
\bibinfo{author}{R.~Ouchene},
\newblock \bibinfo{title}{Numerical simulation and modeling of the hydrodynamic
  forces and torque acting on individual oblate spheroids},
\newblock \bibinfo{journal}{Physics of Fluids} \bibinfo{volume}{32}
  (\bibinfo{year}{2020}) \bibinfo{pages}{073303}.
%Type = Article
\bibitem[{Goniva et~al.(2012)Goniva, Kloss, Deen, Kuipers, and
  Pirker}]{Goniva2012}
\bibinfo{author}{C.~Goniva}, \bibinfo{author}{C.~Kloss}, \bibinfo{author}{N.~G.
  Deen}, \bibinfo{author}{J.~A. Kuipers}, \bibinfo{author}{S.~Pirker},
\newblock \bibinfo{title}{Influence of rolling friction on single spout
  fluidized bed simulation},
\newblock \bibinfo{journal}{Particuology}  (\bibinfo{year}{2012})
  \bibinfo{pages}{582--591}.
%Type = Article
\bibitem[{Moskalensky et~al.(2018)Moskalensky, Yurkin, Muliukov, Litvinenko,
  Nekrasov, Chernyshev, and Maltsev}]{Moskalensky2018}
\bibinfo{author}{A.~E. Moskalensky}, \bibinfo{author}{M.~A. Yurkin},
  \bibinfo{author}{A.~R. Muliukov}, \bibinfo{author}{A.~L. Litvinenko},
  \bibinfo{author}{V.~M. Nekrasov}, \bibinfo{author}{A.~V. Chernyshev},
  \bibinfo{author}{V.~P. Maltsev},
\newblock \bibinfo{title}{Method for the simulation of blood platelet shape and
  its evolution during activation},
\newblock \bibinfo{journal}{PLOS Computational Biology} \bibinfo{volume}{14}
  (\bibinfo{year}{2018}) \bibinfo{pages}{e1005899}.
%Type = Article
\bibitem[{Segr{\'e} and Silberberg(1961)}]{SegreSilberberg1961}
\bibinfo{author}{G.~Segr{\'e}}, \bibinfo{author}{A.~Silberberg},
\newblock \bibinfo{title}{Radial {{Particle Displacements}} in {{Poiseuille
  Flow}} of {{Suspensions}}},
\newblock \bibinfo{journal}{Nature} \bibinfo{volume}{189}
  (\bibinfo{year}{1961}) \bibinfo{pages}{209--210}.
%Type = Article
\bibitem[{McLaughlin(1991)}]{Mclaughlin1991}
\bibinfo{author}{J.~B. McLaughlin},
\newblock \bibinfo{title}{Inertial migration of a small sphere in linear shear
  flows},
\newblock \bibinfo{journal}{Journal of Fluid Mechanics} \bibinfo{volume}{224}
  (\bibinfo{year}{1991}) \bibinfo{pages}{261--274}.
%Type = Article
\bibitem[{Mei(1992)}]{Mei1992}
\bibinfo{author}{R.~Mei},
\newblock \bibinfo{title}{An approximate expression for the shear lift force on
  a spherical particle at finite reynolds number},
\newblock \bibinfo{journal}{International Journal of Multiphase Flow}
  \bibinfo{volume}{18} (\bibinfo{year}{1992}) \bibinfo{pages}{145--147}.
%Type = Article
\bibitem[{Loth and Dorgan(2009)}]{Loth2009}
\bibinfo{author}{E.~Loth}, \bibinfo{author}{A.~J. Dorgan},
\newblock \bibinfo{title}{An equation of motion for particles of finite
  {{Reynolds}} number and size},
\newblock \bibinfo{journal}{Environmental Fluid Mechanics} \bibinfo{volume}{9}
  (\bibinfo{year}{2009}) \bibinfo{pages}{187--206}.
%Type = Article
\bibitem[{Mody and King(2005)}]{Mody2005}
\bibinfo{author}{N.~A. Mody}, \bibinfo{author}{M.~R. King},
\newblock \bibinfo{title}{Three-dimensional simulations of a platelet-shaped
  spheroid near a wall in shear flow},
\newblock \bibinfo{journal}{Physics of Fluids} \bibinfo{volume}{17}
  (\bibinfo{year}{2005}) \bibinfo{pages}{113302}.
%Type = Book
\bibitem[{P{\"o}schel and Schwager(2005)}]{ComputationalGranularDynamics2005}
\bibinfo{author}{T.~P{\"o}schel}, \bibinfo{author}{T.~Schwager},
  \bibinfo{title}{Computational {{Granular Dynamics}}: {{Models}} and
  {{Algorithms}}}, {{SpringerLink B\"ucher}}, \bibinfo{publisher}{Springer
  Berlin Heidelberg}, \bibinfo{address}{Berlin, Heidelberg},
  \bibinfo{year}{2005}.
%Type = Article
\bibitem[{Boyle(1988)}]{Boyle1988}
\bibinfo{author}{J.~Boyle},
\newblock \bibinfo{title}{Microcirculatory hematocrit and blood flow},
\newblock \bibinfo{journal}{Journal of Theoretical Biology}
  \bibinfo{volume}{131} (\bibinfo{year}{1988}) \bibinfo{pages}{223--229}.
%Type = Article
\bibitem[{Yin et~al.(2013)Yin, Thomas, and Zhang}]{Yin2013}
\bibinfo{author}{X.~Yin}, \bibinfo{author}{T.~Thomas},
  \bibinfo{author}{J.~Zhang},
\newblock \bibinfo{title}{Multiple red blood cell flows through microvascular
  bifurcations: {{Cell}} free layer, cell trajectory, and hematocrit
  separation},
\newblock \bibinfo{journal}{Microvascular Research} \bibinfo{volume}{89}
  (\bibinfo{year}{2013}) \bibinfo{pages}{47--56}.
%Type = Article
\bibitem[{Vahidkhah and Bagchi(2015)}]{Vahidkhah2015}
\bibinfo{author}{K.~Vahidkhah}, \bibinfo{author}{P.~Bagchi},
\newblock \bibinfo{title}{Microparticle shape effects on margination, near-wall
  dynamics and adhesion in a three-dimensional simulation of red blood cell
  suspension},
\newblock \bibinfo{journal}{Soft Matter} \bibinfo{volume}{11}
  (\bibinfo{year}{2015}) \bibinfo{pages}{2097--2109}.
%Type = Article
\bibitem[{Zhao et~al.(2012)Zhao, Shaqfeh, and Narsimhan}]{Zhao2012}
\bibinfo{author}{H.~Zhao}, \bibinfo{author}{E.~S.~G. Shaqfeh},
  \bibinfo{author}{V.~Narsimhan},
\newblock \bibinfo{title}{Shear-induced particle migration and margination in a
  cellular suspension},
\newblock \bibinfo{journal}{Physics of Fluids} \bibinfo{volume}{24}
  (\bibinfo{year}{2012}) \bibinfo{pages}{011902}.
%Type = Article
\bibitem[{Zydney and Colton(1988)}]{Zydney1988}
\bibinfo{author}{A.~L. Zydney}, \bibinfo{author}{C.~K. Colton},
\newblock \bibinfo{title}{Augmented solute transport in the shear flow of a
  concentrated suspension},
\newblock \bibinfo{journal}{Physicochemical Hydrodynamics} \bibinfo{volume}{10}
  (\bibinfo{year}{1988}) \bibinfo{pages}{77--96}.
%Type = Article
\bibitem[{Vahidkhah et~al.(2014)Vahidkhah, Diamond, and Bagchi}]{Vahidkhah2014}
\bibinfo{author}{K.~Vahidkhah}, \bibinfo{author}{S.~L. Diamond},
  \bibinfo{author}{P.~Bagchi},
\newblock \bibinfo{title}{Platelet {{Dynamics}} in {{Three-Dimensional
  Simulation}} of {{Whole Blood}}},
\newblock \bibinfo{journal}{Biophysical Journal} \bibinfo{volume}{106}
  (\bibinfo{year}{2014}) \bibinfo{pages}{2529--2540}.

\end{thebibliography}

\end{document}